\documentclass[twocolumn]{aastex631}

\shorttitle{SN~2021wvw: A sub-luminous SN with a short plateau}
\shortauthors{Teja et al.}

\begin{document}

\title{SN~2021wvw: A core-collapse supernova at the sub-luminous, slower, and shorter end of Type IIPs}

\correspondingauthor{Rishabh Singh Teja}
\email{rishabh.teja@iiap.res.in, rsteja001@gmail.com}

\author[0000-0002-0525-0872]{Rishabh Singh Teja}
\affil{Indian Institute of Astrophysics, II Block, Koramangala, Bengaluru-560034, Karnataka, India}
\affil{Pondicherry University, R.V. Nagar, Kalapet, Pondicherry-605014, UT of Puducherry, India}

\author[0000-0003-1012-3031]{Jared A.~Goldberg}
\affil{Center for Computational Astrophysics, Flatiron Institute, 162 5th Ave, New York, NY 10010, USA}

\author[0000-0002-6688-0800]{D. K. Sahu}
\affil{Indian Institute of Astrophysics, II Block, Koramangala, Bengaluru-560034, Karnataka, India}

\author[0000-0003-3533-7183]{G. C. Anupama}
\affil{Indian Institute of Astrophysics, II Block, Koramangala, Bengaluru-560034, Karnataka, India}

\author[0000-0003-2091-622X]{Avinash Singh}
\affiliation{Oskar Klein Centre, Department of Astronomy, Stockholm University, Albanova University Centre, SE-106 91 Stockholm, Sweden}
\affiliation{Hiroshima Astrophysical Science Center, Hiroshima University, Higashi-Hiroshima, Hiroshima 739-8526, Japan}

\author[0000-0002-7942-8477]{Vishwajeet Swain}
\affiliation{Department of Physics, Indian Institute of Technology Bombay, Powai, Mumbai 400076}

\author[0000-0002-6112-7609]{Varun Bhalerao}
\affiliation{Department of Physics, Indian Institute of Technology Bombay, Powai, Mumbai 400076}

\begin{abstract}
We present detailed multi-band photometric and spectroscopic observations and analysis of a rare core-collapse supernova SN~2021wvw, that includes photometric evolution up to 250~d and spectroscopic coverage up to 100~d post-explosion. A unique event that does not fit well within the general trends observed for Type II-P supernovae, SN 2021wvw shows an intermediate luminosity with a short plateau phase of just about 75~d, followed by a very sharp ($\sim$10~d) transition to the tail phase. Even in the velocity space, it lies at a lower velocity compared to a larger Type II sample. The observed peak absolute magnitude is $-16.1$~mag in $r$-band, and the nickel mass is well constrained to $0.020\pm0.006~M_\odot$. Detailed hydrodynamical modeling using \texttt{MESA+STELLA} suggests a radially compact, low-metallicity, high-mass Red Supergiant progenitor ($M_{ZAMS}=18~M_\odot$), which exploded with $\rm \sim 0.2\times10^{51} ~erg~s^{-1}$ leaving an ejecta mass of $M_{ej}\approx5~M_\odot$. Significant late-time fallback during the shock propagation phase is also seen in progenitor+explosion models consistent with the light curve properties. As the faintest short-plateau supernova characterized to date, this event adds to the growing diversity of transitional events between the canonical $\sim$100~d plateau Type IIP and stripped-envelope events. 

\end{abstract}


\section{Introduction} \label{sec:intro}
Stars with mass $\rm\gtrsim8~M_\odot$ end their lives as energetic cosmic explosions called core-collapse supernovae (CCSNe). Depending on factors such as the initial mass, rotation, evolutionary track, retained envelope, immediate surroundings, and explosion energy, their light curves come in various shapes and brightness, and there are differences in their spectral features as well. Persistent hydrogen features in the spectra indicate a Type II class SN; otherwise, a Type I SN \citep{1997Filippenko}. There is a further distinction in the light curve evolution of the Type IIs, with either a linear decline (Type IIL) or with a slow, plateau-like decline (Type IIP) followed by a linear decline  \citep{1979Barbon}. Although appearing distinct, it is becoming increasingly evident that there is no clear boundary between the IIP and IIL subclasses, and these are merely a continuous sequence of the Type II class \citep{2014Anderson, 2019MNRAS.488.4239P}.

Within the Type IIP subclass, there appears to be a certain amount of inhomogeneity concerning the plateau length. In most cases, the plateau length is, on an average, 100~d. However, recent observations have shown several events that deviate from this 100~d plateau length on either side, from the shorter end of the plateau \citep[$\sim4\%$,][]{2018PASA...35...49E, 2021Hiramatsu} to the long plateau phase \citep[$\rm\sim0.35\%~for~>140~d$,][]{2018PASA...35...49E}. A few examples of the short plateau SNe are SN~2006Y (55~d), SN~2006ai (60~d) \citep{2021Hiramatsu}, SN~2020jfo (65~d) \citep{2021A&A...655A.105SJS, 2022ApJ...930...34T}, and SN~2018gj (70~d) \citep{2023ApJ...954..155T}, while SN~2005cs (110~d) \citep{20062005cs, 20092005cs}, SN~2018hwm (130~d) \citep{2021MNRAS.501.1059R}, and SN~2020cxd (120~d), SN~2021aai (140~d), \citep{2022MNRAS.513.4983V} had a longer plateau duration. In addition, there is also a very heterogeneous distribution in the brightness space for this subclass \citep{2016MNRAS.459.3939Valenti}. With more discoveries and extensive follow-up in recent times, many events are found to be intrinsically fainter compared to typical Type II SNe \citep[mean $\rm M_{Vmax}\sim-16.7$,][]{2014Anderson} and are termed as low  ($\rm M_V\geq -15~mag$) or intermediate  ($\rm M_V \approx -16~mag$) luminosity SNe. These low/intermediate luminosity SNe also predominantly show a plateau length of 100~d or more \citep[for instance $\geq140$~d in SN~2016bkv, ][]{2018ApJ...859...78N, 2022MNRAS.513.4983V, 2024arXiv240401776F}. Current understanding attributes these SNe to originate from weak explosions of the lower mass end of the red supergiant (RSG) stars, typically less than 15~$\rm M_\odot$ with low $\rm ^{56}Ni$ mass production \citep{2017MNRAS.464.3013P, 2018MNRAS.473.3863L}.

On the other hand, several short-plateau SNe (50 - 80~d) studies show these to be brighter than the typical Type IIP SNe \citep{2021Hiramatsu, 2022ApJ...930...34T}, with the low-luminosity, short plateau events being infrequent. Type II SNe with short plateau also tend to decline faster during their plateau phase \citep{2021Hiramatsu}. The favored mechanism for these short plateau SNe is still debated. A common trend is that the lightcurve properties require a small but non-negligible H-rich envelope ($\approx$ a few $M_\odot$) at the time of explosion \citep{2021Hiramatsu}. Single-star evolutionary scenarios tend to favor moderate to high initial mass RSGs, as stronger winds in more massive progenitors provide a channel for stars to lose the majority but not the entirety of their H-rich envelope \citep[see, e.g.][]{2010MNRAS.408..827D,2016Sukhbold, 2021Curtis}. Other
theoretical works remain agnostic to the mass-loss mechanism or directly link low envelope mass with binary interaction \citep[e.g.][]{Morozova2015, Paxton2018, Dessart2024}, and a growing body of observational works investigate potential low-mass RSG origins as well \citep{ 2022ApJ...930...34T, 2023ApJ...954..155T, 2024MNRAS.527.6227U} with one being a direct detection \citep{2021A&A...655A.105SJS}. In both progenitor-mass regimes, the envelope mass lost by the progenitor is high. 

The occurrence of the short plateau events is relatively low. While the low luminosity ones might suffer from an observational bias, this cannot be said about most short plateau events since they are usually bright. Even taking binarity into account, population-focused studies \citep[e.g.][]{2018PASA...35...49E,Ercolino2024} nonetheless indicate their rates are expected to be low ($\sim 4\%$).

\begin{figure}[htb!]
    \centering
    \resizebox{\hsize}{!}{\includegraphics{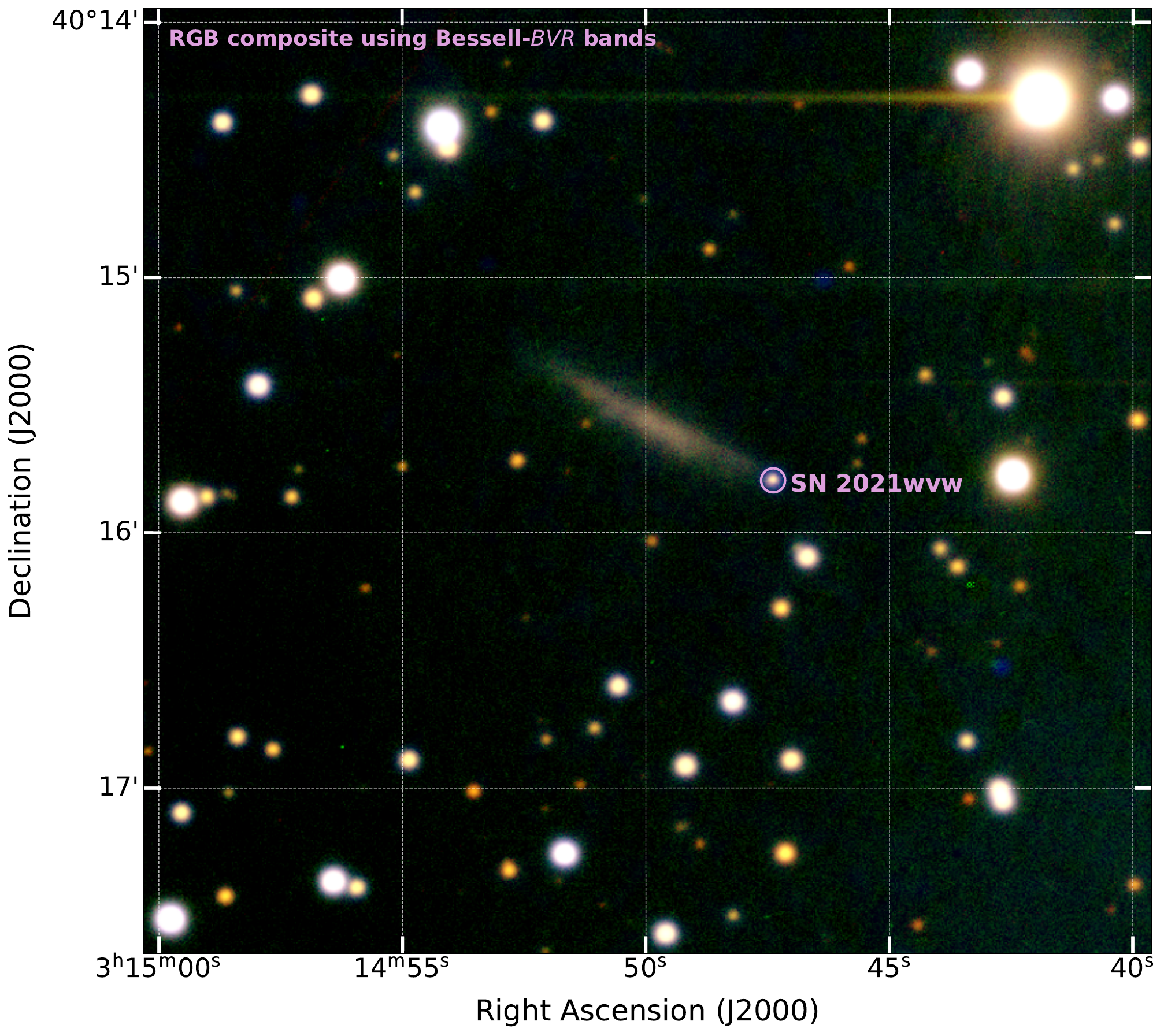}}
    \caption{RGB color composite finder chart for SN2021wvw utilizing Bessell-$BVR$ filters from HCT.}
    \label{fig:finder}
\end{figure}

This work presents comprehensive optical spectroscopic and photometric observations of SN~2021wvw, a distinctive short plateau Type IIP supernova. SN~2021wvw (other names: PS21jnb, ZTF21abvcxel, ATLAS21bgtz, Gaia21eqm) was discovered on August 24, 2021 14:32.6UT (JD=2459451.1) in UGC 02605 \citep{2021TNSTR2917....1JDiscovery} with 17.93 ABMag in the $i-P1$ filter. Subsequently, it was classified as Type II with a strong blue continuum having P-Cygni H$\alpha$ and H$\beta$ emissions \citep{2021TNSCR2941....1HClassification}. The first detection in ZTF-$g$ filter (19.34~mag) was on JD 2459449.95 and the last non-detection in ZTF-$r$ filter (19.15~mag) was on JD 2459449.91. Using this, we obtain JD $2459449.93\pm0.02$ as the explosion epoch. A similar epoch, shifted by $+$0.2~d, is obtained using data from ATLAS forced photometry server with 5-$\sigma$ last non-detection ($>$18.89~mag) on JD 2459449.1 and first detection (18.10$\pm$0.08~mag) on JD 2459451.1 both in ATLAS-$o$ filter. The non-detections in both ZTF-$r$ and ATLAS-$o$ are at a similar epoch, hence we consider this as the last non-detection, and the first detection in ZTF-$g$ band. Using this we obtain $t_{exp}=2459449.9\pm0.3$ as the explosion epoch and use this throughout. The location of SN~2021wvw in its host galaxy is marked in Figure~\ref{fig:finder}. 

The structure of this paper is as follows: Section~\ref{sec:data} provides details of various data sources utilized. The light curves and spectra are analyzed and compared in Section~\ref{sec:analysis} and~\ref{sec:spectra} respectively along with estimating $\rm ^{56}Ni$ mass and expansion velocities. Section~\ref{sec:progenitor} explores the probable progenitor using various models, including complete hydrodynamical modeling, followed by a general discussion in Section~\ref{sec:discussion}. Eventually, we summarize this work in Section~\ref{sec:summary}.  

\section{Photometry and Spectroscopy: Data sources}
\label{sec:data}

\begin{table}[hbt!]
\centering
\caption{Log of spectroscopic observations of SN~2021wvw obtained from HCT.} 
\begin{tabular}{cccc} \hline
   Date          & JD         &Phase$^\dagger$    & Range       \\        
(yyyy-mm-dd)     & (2459000+) &(d)      & (\AA)       \\        
\hline
2021-08-28	&	455.4	&	5.5	&	4000-7700	\\
2021-09-13	&	471.3	&	21.4	&	4000-8900	\\
2021-09-18	&	476.3	&	26.3	&	4000-8900	\\
2021-09-19	&	477.4	&	27.4	&	4000-7700	\\
2021-09-29	&	487.3	&	37.3	&	4000-8900	\\
2021-10-02	&	490.2	&	40.3	&	4000-8900	\\
2021-10-09	&	497.2	&	47.3	&	4000-8900	\\
2021-10-10	&	498.2	&	48.3	&	4000-8900	\\
2021-10-14	&	502.2	&	52.2	&	4000-8900	\\
2021-10-19	&	507.2	&	57.3	&	4000-8900	\\
2021-10-21	&	509.1	&	59.2	&	4000-8900	\\
2021-10-22	&	510.2	&	60.3	&	5300-8900	\\
2021-10-26	&	514.2	&	64.2	&	4000-8900	\\
2021-10-30	&	518.2	&	68.3	&	4000-8900	\\
2021-11-08	&	527.1	&	77.2	&	4000-8900	\\
2021-11-15	&	534.3	&	84.4	&	4000-8900	\\
2021-11-26	&	545.1	&	95.1	&	4000-7700	\\
\hline
\end{tabular} \\ 
{\footnotesize {$^\dagger$Phase given for $t_{exp}=2459449.9$~JD}}
\label{tab:HCTspec}
\end{table}

We began photometry of SN~2021wvw in the optical since +8.4~d past explosion using the 0.7-m robotic GROWTH-India Telescope \citep[GIT,][]{2022AJ....164...90K} and the 2.0-m Himalayan Chandra Telescope (HCT), both situated at the Indian Astronomical Observatory \citep[IAO,][]{2014Prabhu}, Hanle, India. GIT covered dense multi-band photometry in SDSS-$g'r'i'z'$ filters, and HCT covered photometry in Bessell-$V$ and -$R$ filters. We supplemented our observations with photometry from the Asteroid Terrestrial-impact Last Alert System \citep[ATLAS,][]{2018PASP..130f4505T, 2020PASP..132h5002S} forced photometry server \citep{2021TNSAN...7....1S} in $c$ and $o$ filters. We also obtained ZTF \citep{2019ZTF} -$g$ and -$r$ filter apparent magnitudes from ALeRCE \citep{2021AJ....161..242F}. The ATLAS photometry, being noisy, has been binned for 2~d intervals in the late phase using \citet{Youngfp} \texttt{python} script. During the late phase, we took multiple exposures using GIT and HCT and summed them for a better signal-to-noise ratio in respective filters. The SN being far away from the host nucleus ($\sim$ 31$''$) and at the periphery, we do not perform any template subtraction. The last detected photometric points are significantly brighter (1.5-3~mag) than the SDSS photometry\footnote{\url{https://skyserver.sdss.org/dr18/}} in the regions around the host center and near the SN position. Standard photometric data reduction procedures have been adopted utilizing \texttt{IRAF} and \texttt{pyraf}, the details of which can be found in \citet{2023ApJ...954..155T}. The photometric data are tabulated in the Appendix of this work.

We obtained low-resolution ($\rm R\sim800$) optical spectra with the HFOSC instrument available on HCT using 167l slit (1.$\arcsec$92 width and 11$\arcmin$ length). The spectra observed with grisms Gr7 and Gr8 were combined to obtain spectra covering a wavelength range of 4000 to 9000~\AA.
The optical spectra were obtained during 5 - 95~d post-explosion. Beyond 95~d, the SN faded considerably, and spectroscopy with HCT was not feasible. The observed 2-D spectra were bias corrected using nightly bias frames, and the 1-D spectra were optimally extracted. The wavelength correction was performed using the dispersion solutions obtained from several arc lamps (FeNe, FeAr) spectra. The night-sky emission lines (5577, 6300, 6363~\AA) in the background spectra were used to perform the accurate wavelength calibrations, applying small shifts wherever required. Spectrophotometric standards were observed periodically to correct the instrumental response and finalize the spectra in the flux scale. Eventually, a single flux-calibrated spectrum was obtained after combining spectra from individual grisms. All these steps were performed using various tasks in \texttt{IRAF}. 

The host redshift (z=0.0099, \citet{1992ApJS...81....5S}) and line of sight extinction ($E(B-V)=0.24$~mag, \citet{2011ApJ...737..103Sirsa}) are taken from NED and IRSA, respectively. The redshift corresponds to a distance of $41.51\pm 2.91$~Mpc or $\mu = 33.09 \pm0.15$~mag, assuming the $\Lambda CDM$ cosmology with H$\rm_0=72.5~km~s^{-1}~Mpc^{-1}$ \citep{2022ApJ...934L...7R}. We used \citet{Cardelli} extinction law with $R_V=3.1$ to correct for Galactic reddening. We do not find any discernible \ion{Na}{1D} features at redshift of the host galaxy in SN spectra, and hence assume no extinction due to the host galaxy. 

\section{Light Curve Evolution}
\label{sec:analysis}

\begin{figure}[htb!]
    \centering
    \resizebox{\hsize}{!}{\includegraphics{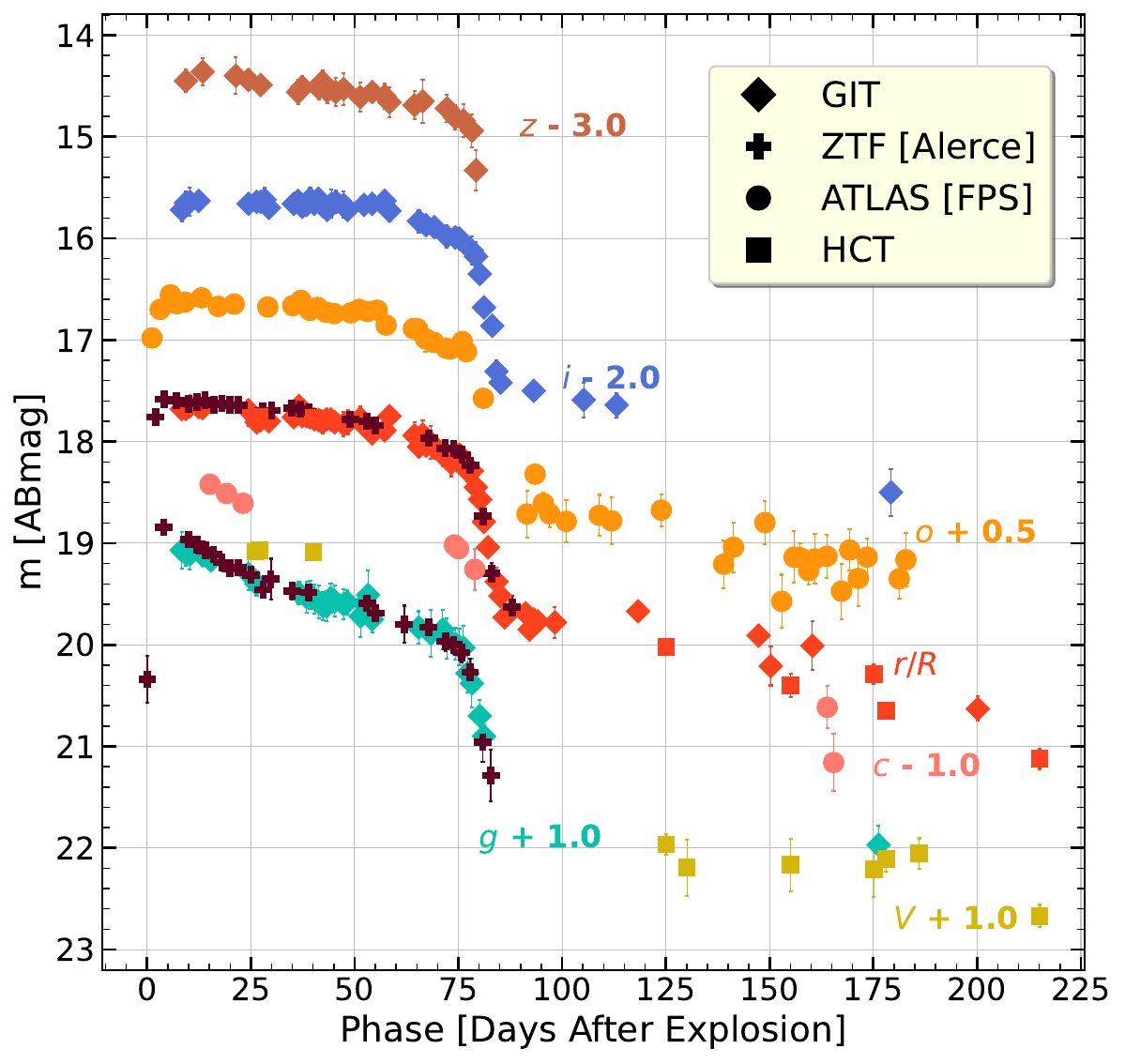}}
    \caption{Light curve evolution of SN~2021wvw for various filters from GIT and HCT is shown. The light curves also include data from ZTF and ATLAS surveys. The constants added to the individual light curves are for visual clarity. }
    \label{fig:photometryevolution}
\end{figure}

We present the panchromatic light curve evolution of SN~2021wvw in Figure~\ref{fig:photometryevolution}. The light curve evolution spans roughly 220~d post-explosion. Other than the bluer bands such as $g$-band, the light curves evolution in different filters show a very flat evolution up to 70 to 80~d before transitioning sharply into the tail phase. The plateau and transition phases are very densely sampled in most filters. In $R$ and $V$ filters, the tail phase is sampled up to 220~d. We estimate a plateau length of around 75~d \citep[OPTd, ][]{2014Anderson} and a sharp transition period of about 10~d. 

\begin{figure}[htb!]
    \centering
    \resizebox{\hsize}{!}{\includegraphics{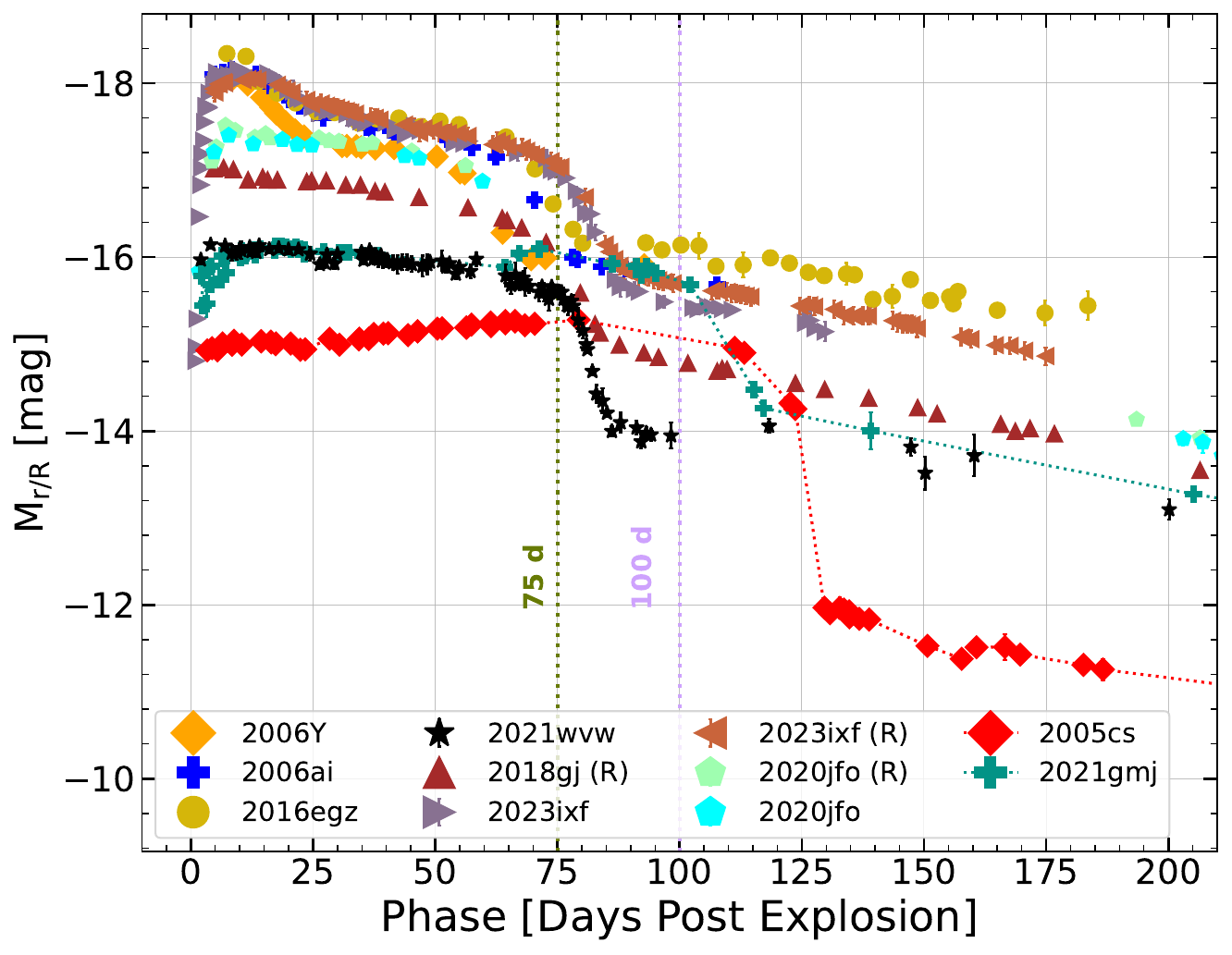}}
    \caption{SN~2021wvw $r$ band light curve evolution is compared with the $r/R$ band light curves of other short plateau SNe. We also show the archetypal low-luminosity SN~2005cs and an intermediate-luminosity SN~2021gmj.}
    \label{fig:lclowlum}
\end{figure}

The mid-plateau absolute magnitude is $\rm \approx -16.0\pm0.1~mag$ in $r$-band. It puts SN~2021wvw in the intermediate luminosity regime for Type IIP SNe. The duration of the plateau phase is also shorter ($\sim75$~d), whereas the typical plateau lengths for Type IIP SNe are $\sim$100~d and even longer in the case of under-luminous SNe (SN~2005cs, SN~2016bkv, SN~2021gmj). In Figure~\ref{fig:lclowlum}, SN~2021wvw $r$-band light curve is compared with $r/R$-band light curves of other intermediate/low luminosity and short plateau SNe, respectively. We compare with the archetypal SN~2005cs \citep{Pastorello2006_2005cs} and SN~2021gmj \citep{2024MNRAS.528.4209M} for low luminosity SNe. Although short plateaus are very rare in the overall Type II SNe, we compare with other well-studied short plateau SNe in literature such as SN 2006Y, SN~2006ai, SN~2016egz \citep{2021Hiramatsu}, SN~2018gj \citep{2023ApJ...954..155T}, SN~2020jfo \citep{2022ApJ...930...34T} and SN~2023ixf \citet{2023ApJ...954L..12T, 2024arXiv240520989S}. 

\begin{table*}[htb!]
    \centering
        \caption{Various slopes obtained for different phases of light curves are presented. Slopes and absolute magnitude for other SNe are also compared. The absolute magnitudes ($M_{r/R}$) are reported from the middle of the plateau.}
    \begin{tabular}{l|cc|ccc|cc|c} \hline
        SN & \multicolumn{2}{|c|}{$g\ [mag~(100~d)^{-1}]$} &\multicolumn{3}{|c|}{$r/R\ [mag~(100~d)^{-1}]$}  &\multicolumn{2}{|c|}{$i\ [mag~(100~d)^{-1}]$} & $M_{r/R}$ \\
        \hline
         & s1&s2 &s1 & s2 & s3 &s2 &s3 &[mag]\\
         \hline
   2021wvw & $2.52\pm0.53$ & $ 1.25\pm0.16$ &$0.34\pm0.19$ &$0.78\pm0.17$  &$0.64\pm0.28$ &$0.10\pm0.13$  &$1.09\pm0.25$ &$-16.0$ \\
    2005cs& - & -&- & $-0.53\pm0.01$ & $0.60\pm0.02$ &-  & - &$-15.2$\\
    2006Y&$4.62\pm0.51$  & $3.28\pm0.10$ &$4.75\pm0.13$ &$0.29\pm0.20$  &-  &$1.22\pm0.10$   &- &$-17.3$\\
    2006ai&  $4.44\pm0.05$& $2.86\pm0.06$ &$4.01\pm0.10$ &$0.96\pm0.04$  &$1.03\pm0.13$  &$1.01\pm0.04$  &$1.53\pm0.19$& $-17.5$ \\
    2016egz&  -& $2.83\pm0.19$ & -& $0.89\pm0.13$ & $1.11\pm0.04$  &$1.51\pm0.17$   &$1.10\pm0.10$ &$-17.6$\\
    2021gmj& - & - &- &$0.25\pm0.01$  &$0.51\pm0.11$  &$1.28\pm0.02$  &$1.27\pm0.03$ & $-15.9$\\
    \hline
    \end{tabular}
    \label{tab:slopes}
\end{table*}

The photospheric phase light curve evolution of SN~2021wvw, particularly for $r$- and $i$- bands, is gradual, which is atypical for short plateau SNe, for which the decline is generally rapid \citep{2021Hiramatsu}. Although the early ($s1$) phase after maximum is not evident in the multi-band light curves, upon closer inspection, the $g$- and $r$- bands show a gradual decline in the post-peak evolution. We find this to be $\rm 2.52^{+0.52}_{-0.53}~mag~100~d^{-1}$ and   $\rm 0.34^{+0.19}_{-0.19}~mag~100~d^{-1}$ in $g$- and $r$- band, respectively, whereas the decline is much steeper in other objects: for example, SN~2006Y and SN~2006ai have  $\rm 4.62^{+0.51}_{-0.52}~mag~100~d^{-1}$ and $\rm 4.44^{+0.05}_{-0.05}~mag~100~d^{-1}$ respectively in $g$-band. We also estimated the decline rates of the plateau phase (s2) and tail phase (s3). The estimated values of various slopes and the mid-plateau absolute magnitudes are shown in Table~\ref{tab:slopes}. Interestingly, the plateau phase in the $i-$ band for SN~2021wvw is almost non-declining with $\rm s2=0.10\pm0.13~mag~100~d^{-1}$, whereas for other SNe, both with lower luminosity and shorter-plateau SNe, it is around an order of magnitude higher. Evidently, the tail phase decline ($\rm s3=0.64\pm0.28mag~100~d^{-1}$) of SN~2021wvw in the $r$-band is close to the values obtained for other lower luminosity SNe (SN~2021gmj, SN~2005cs). At a similar phase, slope s3 in the $i$-band is non-differentiable for both low-luminosity and short-plateau SNe with values ranging from 1.1 to 1.5~mag$\rm~100~d^{-1}$. 

Comparing the mid-plateau luminosity ($\rm M_{tp1/2}$) in $r/R$-band with other SNe, we find that SN~2021wvw has a similar magnitude as of SN~2021gmj ($\rm -15.9~mag$), and about $\rm 1~mag$ higher than the $\rm M_{tp1/2}$ of SN~2005cs ($\rm -15.2~mag$). $\rm M_{tp1/2}$ of a majority of other short-plateau SNe is higher than $\rm -17~mag$ except for SN~2018gj ($\rm -16.7~mag$) as shown in Figure~\ref{fig:lclowlum}.

\subsection{Radioactive \texorpdfstring{$\rm ^{56}Ni$}{}}

The late-time evolution is primarily powered by the radioactive decay of  $\rm ^{56}Ni$ formed during explosive nucleosynthesis. It is the ultimate powering source in the Type II SNe during the nebular phase. Hence, the late-time bolometric light curve can provide tight constraints on estimating the $\rm ^{56}Ni$ mass.
We use \texttt{SuperBol} \citep{2018RNAAS...2..230N} to estimate the pseudo-bolometric light curve and a complete bolometric light curve evolution from extrapolated blackbody estimates. The extinction-corrected multi-band light curves were used as input, taking well-sampled $r$-band as the reference light curve. The filters utilized for the pseudo-bolometric curves were $groiz$.

We estimate the $\rm ^{56}Ni$ using the following equation given in \citet{2016MNRAS.461.2003Y} which also takes into account the $\gamma-$ray leakage in case of an stripped envelope:
$$
    \label{eq:Nitc}
    L_{obs}(t) = L_0\times M_{Ni}\times\left [ e^{-(\frac{\Delta t}{t_{Co}})}-e^{-(\frac{\Delta t}{t_{Ni}})}\right ]\times \left ( 1-e^{(-\frac{t_c^2}{(\Delta t)^2})}\right)
$$
where, $\Delta t=t-t_{exp}$, $\rm L_0$, $\rm t_{Co}$, and $\rm t_{Ni}$ are $\rm 1.41\times10^{43}\ erg~s^{-1}$, $\rm 111.4~d$ and $\rm 8.8~d$, respectively. Here, $t_c$ is the characteristic time when the optical depth for $\gamma-$rays approaches unity \citep{2016MNRAS.461.2003Y}. We use \texttt{scipy} and \texttt{emcee} packages to fit and estimate errors in the values. Using the pseudo-bolometric light curve, we estimate $\rm ^{56}Ni$ mass as $\rm 0.011\pm0.001~M_\odot$. This provides a lower limit on the $\rm ^{56}Ni$ mass. In addition, considering the blackbody fitted luminosity as bolometric luminosity, we obtain $\rm M_{^{56}Ni} =\rm 0.023\pm0.003~M_\odot$, which we consider as an upper limit for the estimated values. The latter value is more than twice what was obtained using the pseudo-bolometric light curve, but, synonymous with a $\sim$\,50\% NIR contribution seen in the nebular phase of SN~2023ixf \citep{2024arXiv240520989S}. We lack NIR data to provide more information about the accuracy of the contribution in the late phase. Nevertheless, the $^{56}$Ni mass estimated implies a significant NIR flux contribution at late phases. Hence, NIR observations for such objects in the nebular phase are crucial for a better understanding. In subsequent sections, we perform light curve modeling to constrain the nickel mass and other parameters more robustly.  

\section{Spectra}
\label{sec:spectra}
\begin{figure}[htb!]
    \centering
    \resizebox{\hsize}{!}{\includegraphics{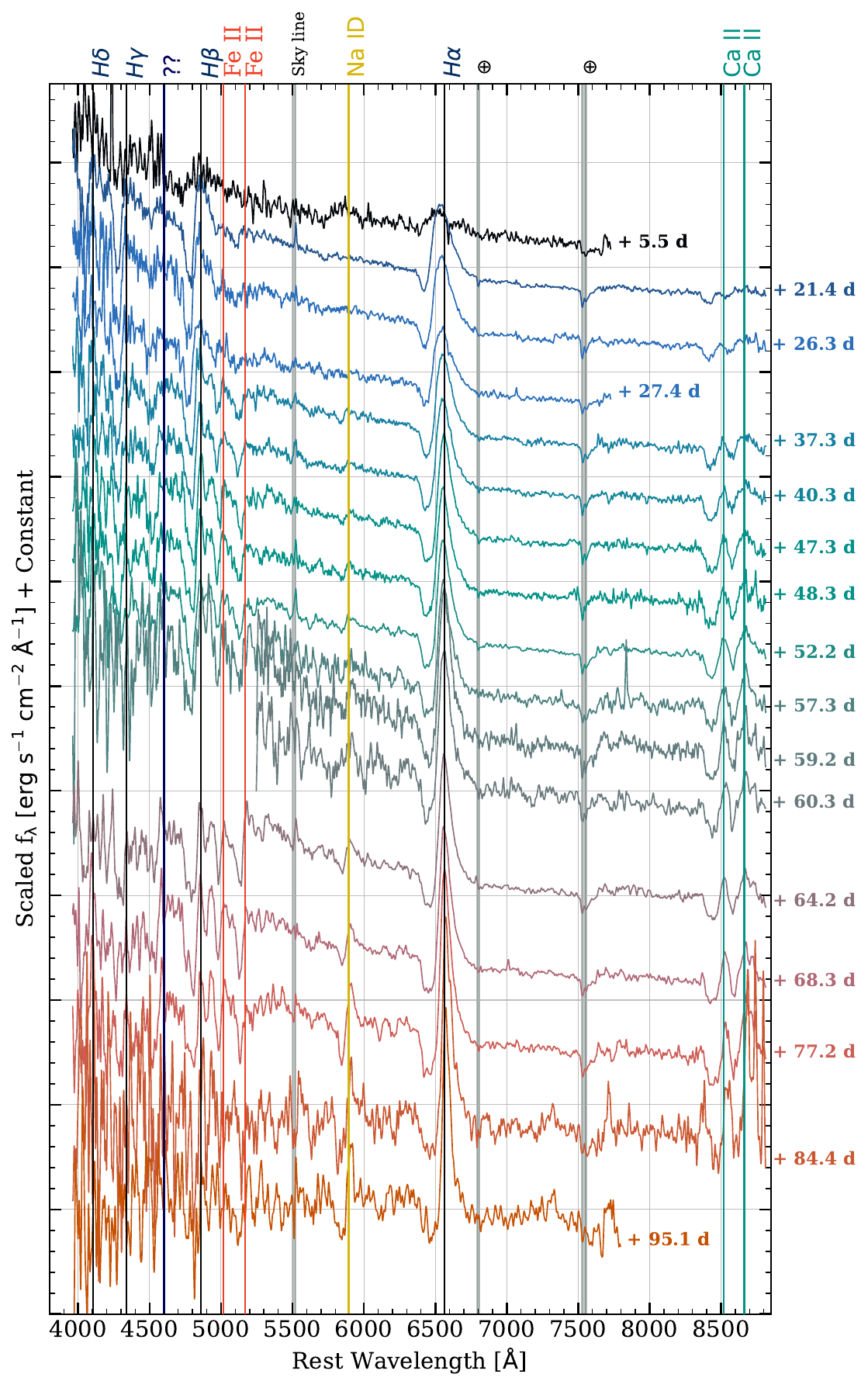}}
    \caption{Spectral sequence for SN~2021wvw. The spectra have been corrected for absolute flux using corresponding photometry and also de-reddened using MW LOS extinction.}
    \label{fig:spectraevolution}
\end{figure}

We present a complete spectral evolution of SN~2021wvw covering the plateau and transition phases in Figure~\ref{fig:spectraevolution}. The spectra have been calibrated with the corresponding multi-band fluxes, corrected for the host redshift, and de-reddened with the estimated extinction. The phases mentioned are with respect to the estimated explosion epoch. All the well-identified lines are marked for clarity in the figure. 

\begin{figure}[htb!]
    \centering
    \resizebox{\hsize}{!}{\includegraphics{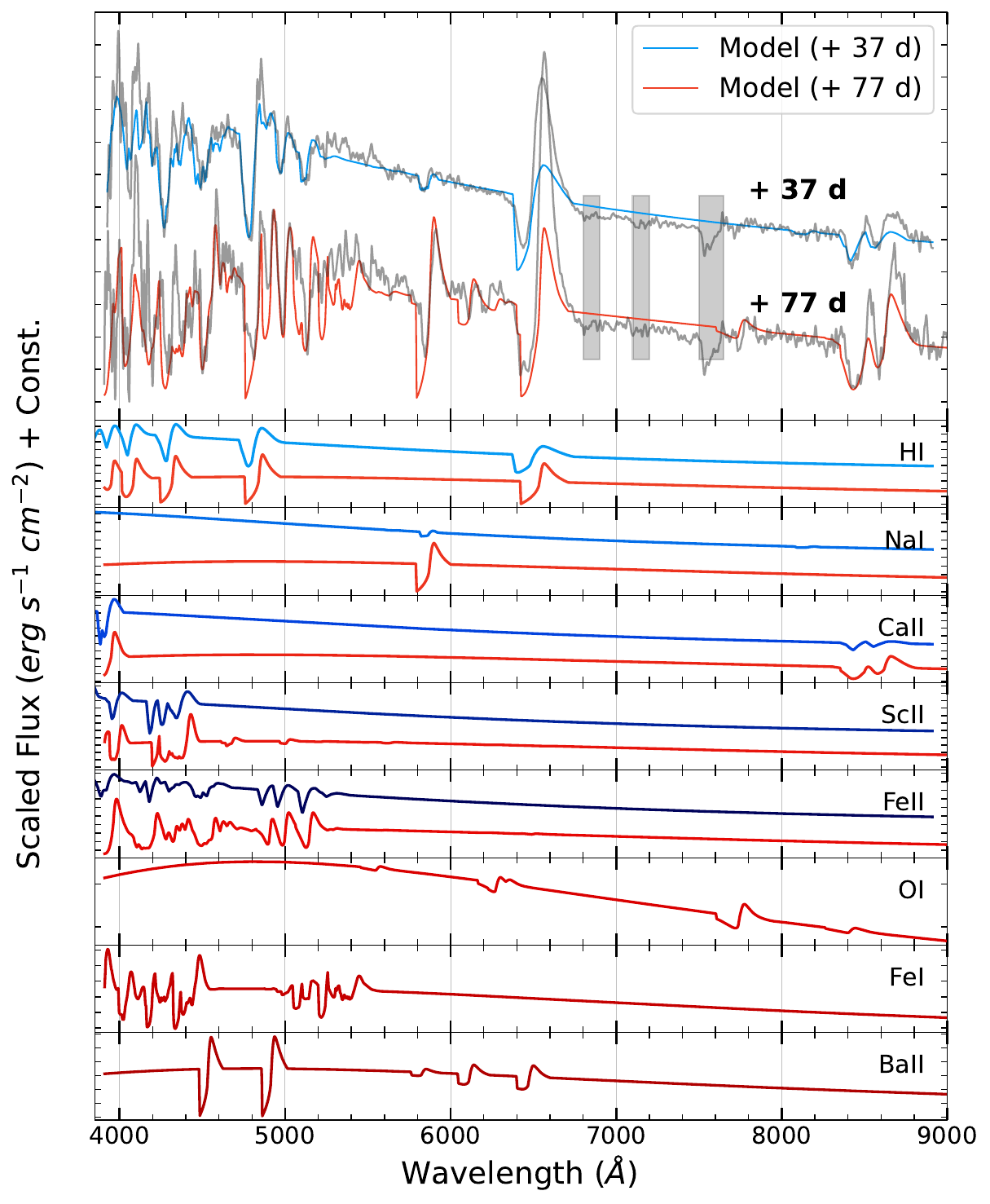}}
    \caption{SYNAPPS model fitting to the observed spectra around the mid and end plateau phases. The lower small panels show the model spectra of individual species when the contribution from rest of the species is turned off. }
    \label{fig:syn1}
\end{figure}
\subsection{Evolution and comparisons}

The first spectrum was obtained at $\rm +5.5~d$. The spectrum comprises a blue continuum with broad Balmer features and \ion{He}{1} $\lambda~5876$ superposed on it. After that, there is a gap of around 15~d; the following spectrum is on +21~d. Thereafter, the spectral evolution is densely sampled until the supernova enters the nebular phase. Qualitatively, absorption features appear relatively narrow at first glance compared to typical IIP SNe, indicating relatively low velocities, and they become narrower with time. Although there are some hints of \ion{Fe}{2} in the bluer region at $\rm+21~d$, these features do not evolve much till $\rm+37~d$, after which we start to see \ion{Fe}{2} lines conspicuously. Around $\rm+21~d$,  we also observe the appearance of \ion{Ca}{0} NIR triplet feature in the red-ward region, strengthening as the SN ejecta evolves further in the photospheric phase. Interestingly, the region between \ion{Fe}{2} lines and H$\alpha$ is devoid of any lines except a weak appearance of \ion{Na}{1D} from $\rm+37~d$ onward. Similarly, the region between H$\alpha$ and \ion{Ca}{0} triplet lacks any discernible features until the end of the observed evolution. We observe a band of emission lines between H$\beta$ and H$\gamma$, usually attributed to \ion{Fe}{0} lines.

The mid-plateau and end-plateau spectra are modeled using \texttt{SYNAPPS/Syn++} \citep{2011PASP..123..237T} to better ascertain the minimum number of species required to explain the observed spectra. \texttt{SYNAPPS/Syn++} is a direct implementation of parameterized spectral synthesis code \texttt{SYNOW} \citep{2010ascl.soft10055P}. It assumes a spherical symmetry with homologous expansion of the ejecta. The emission of photons is from a sharp photosphere, with the optical depth taken as an exponential function of velocity.
$$ \tau_{ref}(v)=\tau_{ref}(v_{ref})\exp{\left (\frac{v_{ref}-v}{v_e} \right )}$$ where $v_{ref}$ is reference velocity for parameterization and $v_e$ is the maximum velocity allowed at the outer edge of the line-forming region \citep{2011PASP..123..237T}. For a particular optical depth, the reference line profile is estimated for a given ion with the remaining lines following Boltzmann statistics \citep{2010ascl.soft10055P}. \texttt{SYNAPPS} iteratively generates synthetic spectra based on a provided input file with parameters such as ions list, blackbody temperature, expansion velocities, and opacities. The synthetic spectra thus obtained are compared with the observed spectra for each iteration. The procedure is automated and requires only initial input parameters with user-defined ranges for each parameter to constrain the parameter space physically. \texttt{SYNAPPS} has been predominantly used to model stripped-envelope and thermonuclear SNe spectra but has been successfully utilized in a number of hydrogen-rich SNe cases as well \citep{2012MNRAS.419.2783T, 2013MNRAS.433....2S, 2019MNRAS.485.5120B, 2021MNRAS.504.1009D}.

\begin{figure}[htb!]
    \centering
    \resizebox{\hsize}{!}{\includegraphics{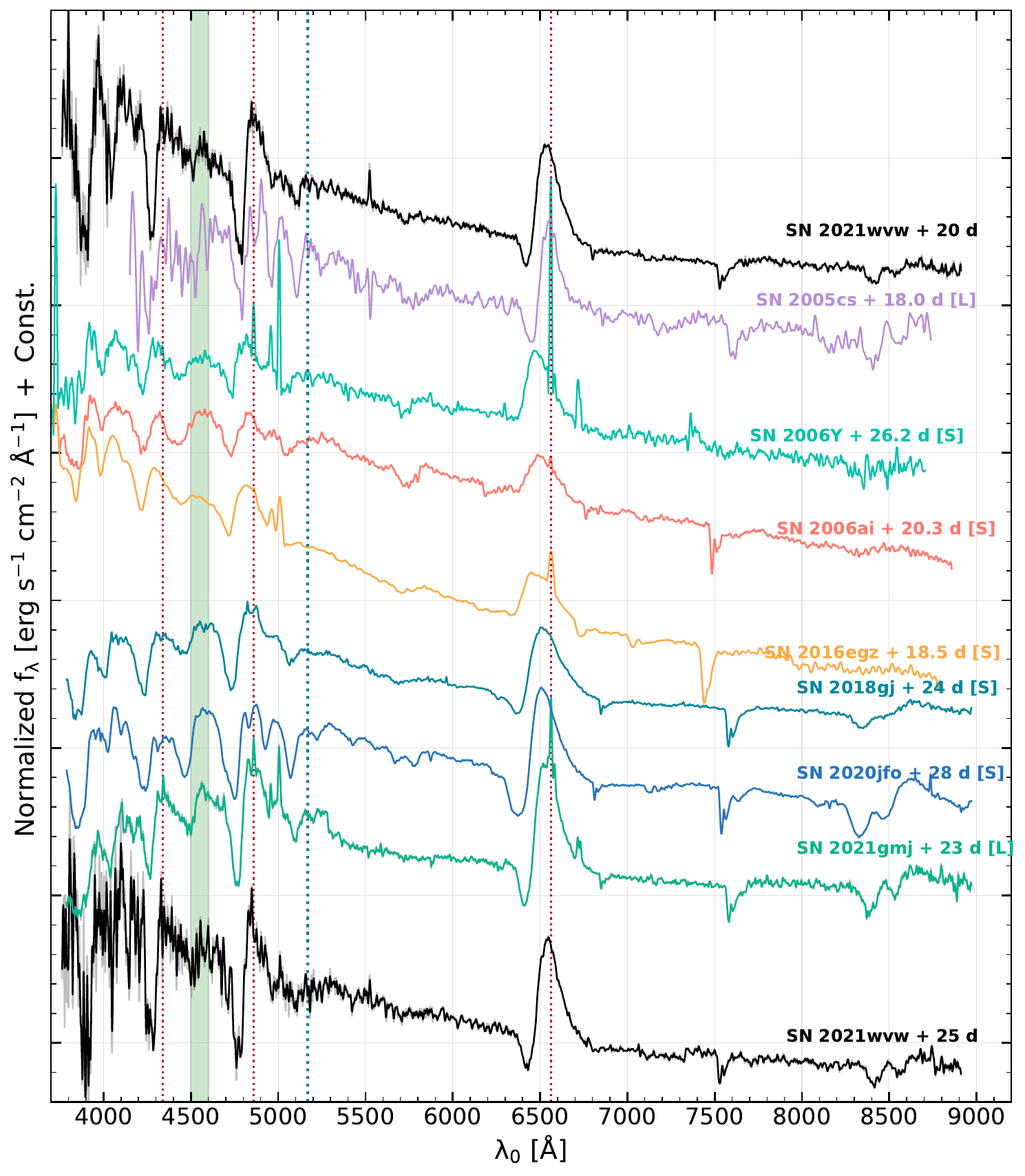}}\\
      \resizebox{\hsize}{!}{\includegraphics{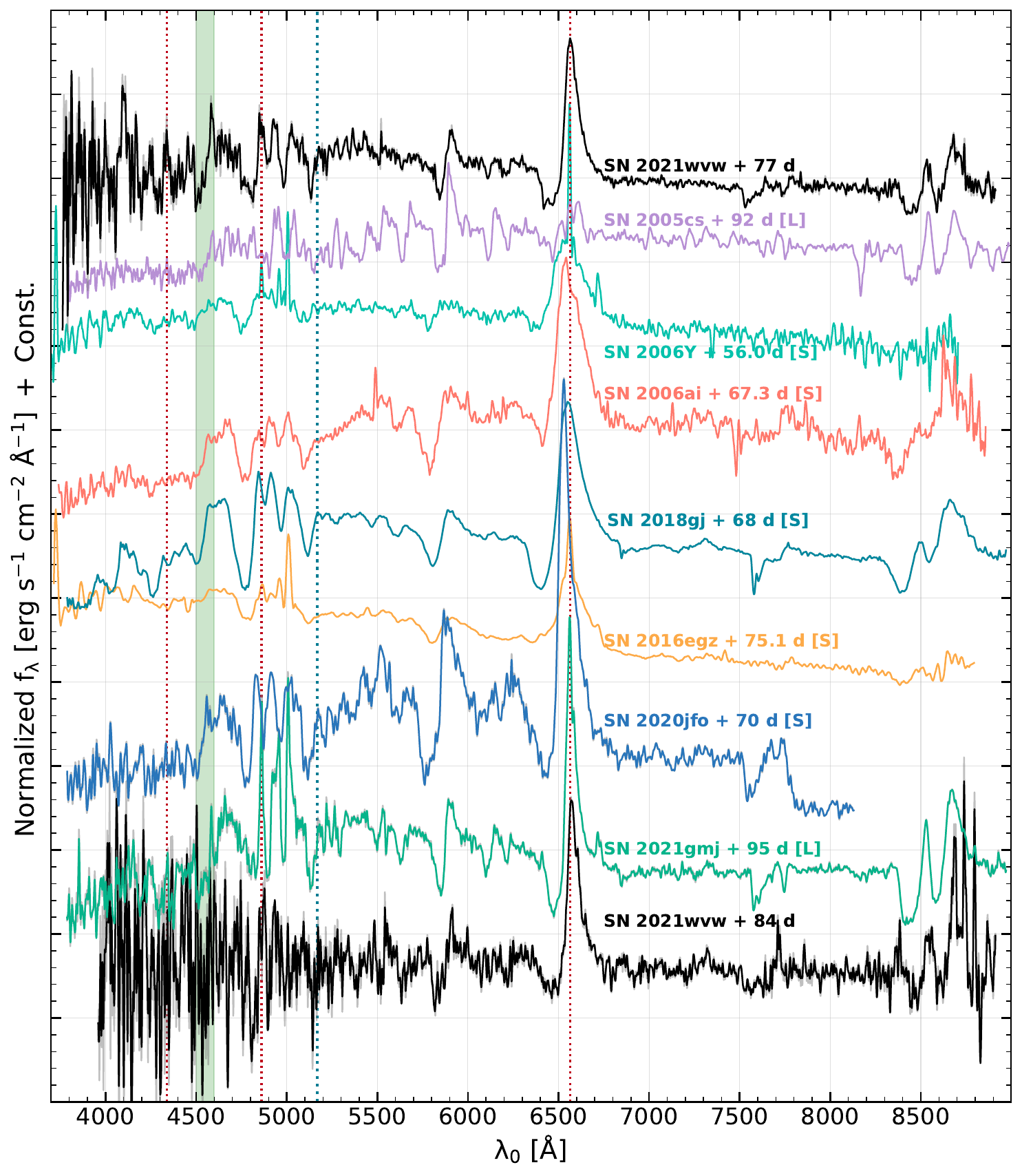}}
    \caption{Spectral comparisons at the early and late plateau phase with short-plateau SNe and with other sub-luminous SNe.}
    \label{fig:speccomp}
\end{figure}

For the first setup to model the $\rm+37~d$ spectrum, we include only five species namely \ion{H}{1}, \ion{Ca}{2}, \ion{Na}{1}, \ion{Sc}{2} and \ion{Fe}{2}. The overall best-fit spectra and various species contributions are shown in Figure~\ref{fig:syn1}. The individual species contributions are obtained by utilizing the best-fit output as input in \texttt{syn++} by turning on one species at a time in the input file. No warping function is applied, i.e., a1=a2=0. Only a0 is varied, which signifies the flux level. Photospheric velocity obtained on the day $\rm+37~d$ is 3830$\rm~km~s^{-1}$. 

For end-plateau spectra at $\rm+77~d$, in addition to the previously included species, we add three more metal species, namely \ion{Ba}{2}, \ion{Fe}{1}, and \ion{O}{1}. Further, we find that the broad emission band around 4800~\AA\ is a blend of multiple metal lines originating from neutral Fe, \ion{Sc}{2}, and \ion{Ba}{2}. The photospheric velocity obtained from these fits around $\rm+77~d$ is 2170$\rm~km~s^{-1}$. 

In Figure~\ref{fig:speccomp}, we compare the SN~2021wvw spectra with a few other short-plateau SNe along with the low-luminosity SN~2005cs and intermediate luminosity SN~2021gmj. Firstly, in the top panel of Figure~\ref{fig:speccomp}, we compare the spectra around 20~d when the metal features are well developed. We clearly observe that there are similarities as well as dissimilarities in the spectral features. At first glance, the features appear similar to SN~2005cs and SN~2021gmj, i.e., narrow and strong absorption. The H$\alpha$ absorption appears shallow, which seems to be the general trend for the short-plateau SNe and is completely indiscernible in some of the brighter and fast-declining short-plateau SNe, for example, SN~2006Y, SN~2006ai, and SN~2016egz. At similar epochs, other SNe have well-developed metal features such as \ion{Fe}{2} lines toward the blue end, whereas we only see a hint of these lines in SN~2021wvw. In the bottom panel of the same figure, we compare SN~2021wvw spectra during the end plateau phase, where we find the appearance of the strongest metallic features. Comparison spectra for other SNe are taken at similar epochs. The SN~2021wvw spectra show similar features to other sub-luminous SNe, but the absorption depths are shallow. However, SN~2021wvw has well-developed P-Cygni (more representative of a typical Type IIP) profiles compared to much shallower absorption depths in short-plateau SNe SN~2006Y, SN~2006ai, and SN~2016egz.

\subsection{Velocities}

\begin{figure}[htb!]
    \centering
    \resizebox{\hsize}{!}{\includegraphics{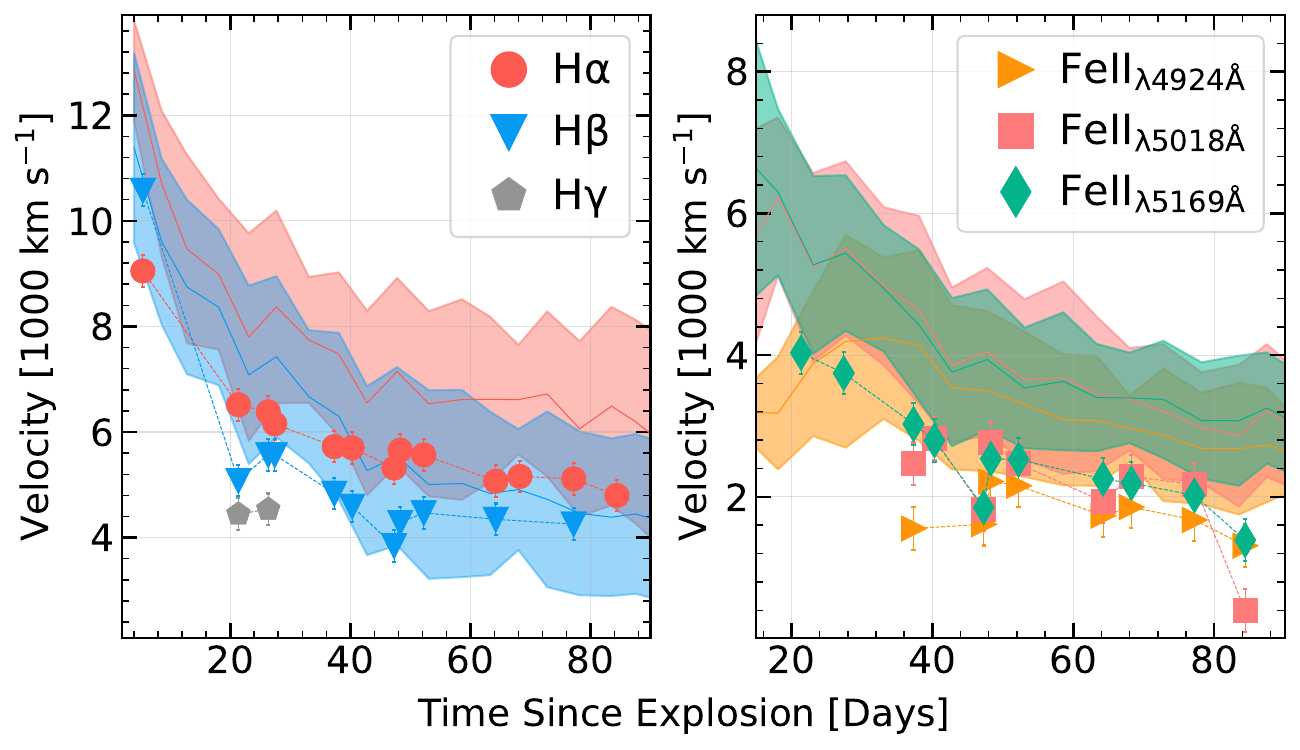}}
    \caption{Expansion velocity evolution estimated from several prominent metallic features (including Balmer lines) observed in the spectra. The velocities have been compared with a large sample taken from \citet{Gutierrez2017_TypeIISample}. The shaded region gives the corresponding 1-$\sigma$ scatter around the sample mean. }
    \label{fig:vel}
\end{figure}

We utilize some of the well-resolved absorption features to estimate expansion velocities of the ejecta. We iteratively measure the absorption minimum of these lines using \texttt{IRAF} by fitting an inverted Gaussian assuming a multitude of continuum points. The absorption minima are corrected for redshift and eventually converted to the expansion velocities using the central rest wavelengths of the corresponding features. We have estimated these velocities for six lines as shown in Figure~\ref{fig:vel}. The errors in velocity estimates are much smaller than the instrumental resolution; hence, the latter has been quoted as the errors in the velocities. 

For the first epoch ($\rm+5.5~d$), we could identify the absorption dips blueward of H$\alpha$ and H$\beta$ rest wavelengths corresponding to $\rm \sim9,100~km~s^{-1}$ and $\rm \sim10,600~km~s^{-1}$ line velocity respectively. At $\rm+21~d$, apart from Balmer features, we could measure the velocity from \ion{Fe}{2}~5169~\AA. Up to $\rm+85~d$, the velocities are measured, and their time evolution is shown in Figure~\ref{fig:vel}. Around $\rm+40~d$, which is proximal to the mid-plateau mark, we measure the H$\alpha$ and \ion{Fe}{2}~5169~\AA\ velocities as $\rm \sim5,700~km~s^{-1}$ and $\rm \sim2,800~km~s^{-1}$, respectively. The \texttt{SYNAPPS} modeling around similar phase gives a value which is between these two values ($\rm \sim3830~km~s^{-1}$). As the ejecta evolves, the expansion velocities keep decreasing until we can confidently resolve the absorption minimum. Towards the end of the plateau phase, around +75~d, we find the expansion velocities to be $\rm \sim2,000~km~s^{-1}$ from \ion{Fe}{0} lines and $\rm \sim5,100~km~s^{-1}$ from H$\alpha$. The model spectrum around a similar phase gives $\rm \sim2,170~km~s^{-1}$ as the photospheric velocity, which is much closer to the values obtained from the metallic features. 

We further compare these velocities with the mean expansion velocities obtained from a larger sample of Type II SNe \citep{Gutierrez2017_TypeIISample}. The mean velocities and 1-$\sigma$ scatter in these are overplotted in Figure~\ref{fig:vel}. We see that the SN~2021wvw velocities lie at the lower 1-$\sigma$ end of the sample, implying that this is a slowly evolving ejecta. For metal lines,  the velocities are even smaller than the lower 1-$\sigma$ edge from the sample. Around mid-plateau, the difference between the mean velocities of the sample and SN~2021wvw observed velocities is $\rm \sim1,500~km~s^{-1}$.  

\section{Plausible Progenitor}
\label{sec:progenitor}
\subsection{Semi-analytical models}
\begin{figure}[htb!]
    \centering
    \resizebox{\hsize}{!}{\includegraphics{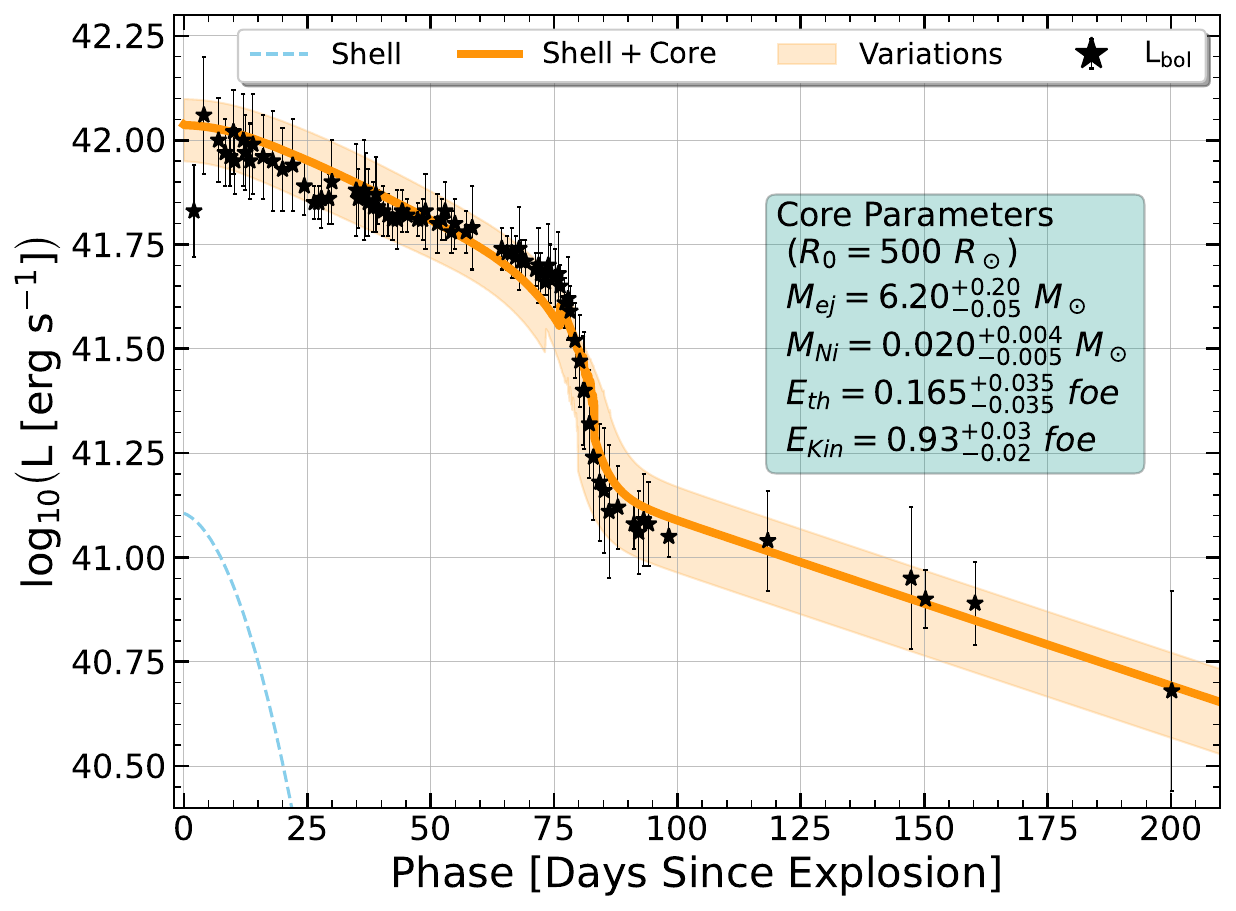}}
    \caption{Semi-analytical fit for fixed radii of 500~R$_\odot$. The values provided in the inset are for the best-matching models.}
    \label{fig:nagy1}
\end{figure}

We attempt to model the bolometric light curve of SN~2021wvw using a two-component progenitor model \citep{NagyVinko2016} to roughly constrain a few parameters and motivate detailed modeling. This formulation comprises of a dense `core' and an extended `envelope,' representing the bulk of the ejecta and the near-surface layers, respectively. \citet{ArnettFu1989} form the basis of this semi-analytical formulation with subsequent additions by \citet{BlinnikovPopov1993, Nagy2014} to obtain approximate ejecta mass ($\rm M_{ej}$), progenitor radius ($\rm R_0$), energy ($\rm E_{tot}$), and $\rm ^{56}Ni$ mass ($\rm M_{Ni}$) estimates. The total energy $\rm E_{tot}$ comprises ejecta kinetic energy ($\rm E_k$) and initial thermal energy ($\rm E_{th}$) deposited by the shock ($\mathrm{E_{tot}=E_k+E_{sh}}$). In this analytical formulation, the SN ejecta, which is spherically symmetric, is divided into two components: a) an interior core with a constant (or flat) density; b) an outer less dense shell with an exponential (n=2) density profile \citep{NagyVinko2016}. These components have independent sets of physical parameters, with the origin of the radius being the same. The contribution from each component to the light curve is estimated independently.  

There is a degeneracy among various parameters \citep{NagyVinko2016}. In a similar analysis for two other short plateau SNe, SN~2018gj and SN~2020jfo, the progenitor radii did not match well with the results obtained using detailed hydrodynamical modeling \citep{2023ApJ...954..155T, 2022ApJ...930...34T}. So, in this work, we do not attempt to constrain the radius of the progenitor; instead, we fix the radius to multiple values beforehand. We take three cases: a fairly compact progenitor (300~$\rm R_\odot$), a typical RSG radius (700~$\rm R_\odot$), and a radius in between (500~$\rm R_\odot$). We vary other parameters to get a light curve matching the observed light curve. Another caveat to consider is the lack of early UV and $U$-band data, which, in models, is usually governed by the shell part. This outer envelope could also act as proximal CSM around the RSG progenitor \citep{NagyVinko2016}. Due to lack of data, no attempts were made to estimate CSM. Instead, we fixed the shell values (to a negligible contribution) so that they do not affect the early light curve. Since the models are analytical, the errors are estimated by first obtaining a match to the observed light curve data, followed by varying the parameters to fit the upper and lower error bars associated with the observed light curve. The best parameters obtained for the fixed radii values are presented in Table~\ref{tab:nagy}. We could find that the model fits equally well with very similar parameters within error bars for each radii value. The case for 500~$\rm R_\odot$ is shown in Figure~\ref{fig:nagy1}. 

The best-fit values of the parameters $\rm M_{ej}$ and $\rm M_{Ni}$ do not vary much for the different radii considered here. The only considerable changes are in the energy values. From these models, we find the $\rm M_{ej}$ to be $\rm \sim 6.5~M_\odot$, $\rm M_{Ni} = 0.020\pm0.005~M_\odot$, and a total energy between 1.1 to 1.3 foe. The $\rm M_{ej}$ values for SN~2021wvw are similar to those obtained in other short plateau cases (for example, SN~2018gj, SN~2020jfo) but with lower explosion energy. The lower energy values are expected for SN~2021wvw, considering its sub-luminous nature. The total energy contribution from the core in the case of low-luminosity SN~2005cs is $\rm\sim0.5~foe$ \citep{NagyVinko2016} with $\rm M_{ej}=8.0~M_\odot$. Considering the intermediate brightness of SN~2021wvw and a shorter plateau length, the estimated parameters are reasonably well constrained with tight bounds on the $\rm ^{56}Ni$ mass. Using these values as our reference point, we delve into more details about the progenitor and its origins using complete hydrodynamical modeling. 

\begin{table}
\centering
\caption{Core parameters for best matching semi-analytical models} 
\begin{tabular}{l|c|c|c} \hline
   Parameters$^*$          & R=300~$\rm M_\odot$        & R=500~$\rm M_\odot$   &    R=700~$\rm M_\odot$       \\        
\hline
$\rm M_{ej}$ ($\rm M_\odot$) & $6.50^{+0.20}_{-0.20}$& $6.20^{+0.20}_{-0.05}$ & $6.60^{+0.10}_{-0.10}$ \\
$\rm E_{th}$ ($\rm 10^{51}~erg$)& $0.27^{+0.13}_{-0.05}$ & $0.17^{+0.04}_{-0.04}$ & $0.12^{+0.03}_{-0.02}$\\
$\rm E_{kin}$ ($\rm 10^{51}~erg$)& $1.00^{+0.20}_{-0.12}$ & $0.93^{+0.03}_{-0.02}$ & $1.05^{+0.01}_{-0.01}$\\
$\rm M_{Ni}$ ($\rm M_\odot$) &$0.020^{+0.004}_{-0.006}$ &$0.020^{+0.004}_{-0.005}$ & $0.020^{+0.004}_{-0.005}$\\
\hline
\end{tabular}\\
$^*$ $\rm T_{rec}\approx6000~K$, $\rm A_g=6.5\times10^{10}~d^2$
\label{tab:nagy}
\end{table}

\subsection{Hydrodynamical Modeling}

In the previous section, we obtained rough estimates of the progenitor parameters. Unfortunately, we lack the nebular phase spectra, which could also be utilized to constrain the progenitor's C/O core mass. Initially, we looked for models representative of SN~2021wvw evolution in other previous studies. However, none of the grids of model light curves or individual models available in the literature could provide a short plateau length with low luminosity \citep{2010MNRAS.408..827D, 2018PASA...35...49E, 2023PASJ...75..634M}. For the case of short-plateau SNe, it has been noticed that a wide range of plausible RSG masses could give rise to these SNe ranging from $\rm 8-12~M_\odot$ \citep{2021A&A...655A.105SJS, 2022ApJ...930...34T, 2024MNRAS.527.6227U} and reaching up to $20-30$~$\rm M_\odot$ \citep{2010MNRAS.408..827D, 2021Hiramatsu}. Therefore, to ascertain the properties of the plausible progenitor of SN~2021wvw, its evolutionary scenario, mass loss before the explosion, explosion energy, and ejecta mass, we perform the hydrodynamical modeling by evolving progenitors for both the lower and higher end of RSGs, allowing arbitrarily enhanced winds to mimic the impact of prior mass loss (due to binary interaction \citealt[e.g.][]{Laplace2021,Ercolino2024} or eruptive mass loss during the star's life \citealt[e.g.][]{Cheng2024}) on the mass of the H-rich ejecta. 

We use the 1-D stellar evolution code \texttt{MESA} \citep{Paxton2011,Paxton2013,Paxton2015,Paxton2018,Paxton2019,Jermyn2023} revision 15140 to evolve the progenitor and hydrodynamical explosion and \texttt{STELLA} \citep{Blinnikov2004,Baklanov2005,Blinnikov2006} to obtain synthetic observables, specifically light curves and expansion velocity. Most of the parameters in MESA were kept the same as provided in the inlists \texttt{make\_pre\_ccsn\_IIp} and \texttt{ccsn\_IIp} and described in detail in \citet{Farmer} and \citet{Paxton2018}; additional detailed descriptions of the setup and key parameters are mentioned in \citep{2022ApJ...930...34T, 2023ApJ...954..155T}. We use the binding-energy fallback scheme introduced in \citet{Paxton2019,2019ApJ...879....3GGOLD} to quantify late-time fallback during the shock propagation phase. In this work, we mainly focus on the following parameters: zero-age main-sequence (ZAMS) mass, metallicity (z), wind scaling factor ($\alpha_{wsf}$),
mixing length ($\alpha_{MLT}$), explosion energy, nickel mass, and explosive mixing via the Duffell Rayleigh Taylor Instability (RTI) \citep{2016ApJ...821...76D} 1D implementation by varying the ratio of RTI parameter $\eta_{R,e}$ and diffusion parameter $\eta_R$ \citep{Paxton2018}. The progenitor models are exploded in \texttt{MESA} via a thermal energy injection to a specified total
explosion energy, and the ejecta evolution is followed to just before shock breakout following \citet{Paxton2018} as discussed in \citet{2022ApJ...930...34T, 2023ApJ...954..155T} making use of the \citet{2016ApJ...821...76D}
implementation for mixing via the Rayleigh-Taylor Instability. The models are then handed off to \texttt{STELLA} when the shock reaches an overhead mass coordinate of 0.05~$\rm M_\odot$. Before evolving a new set of progenitors, we first try the short-plateau models from previous works, namely SN~2020jfo \citep{2022ApJ...930...34T} and SN~2018gj \citep{2023ApJ...954..155T}. Exploding these with lower energies to match the plateau luminosities makes the plateau length longer, leaving these models infructuous. We then proceed to evolve additional models. 

\begin{figure}[htb!]
    \centering
    \resizebox{\hsize}{!}{\includegraphics{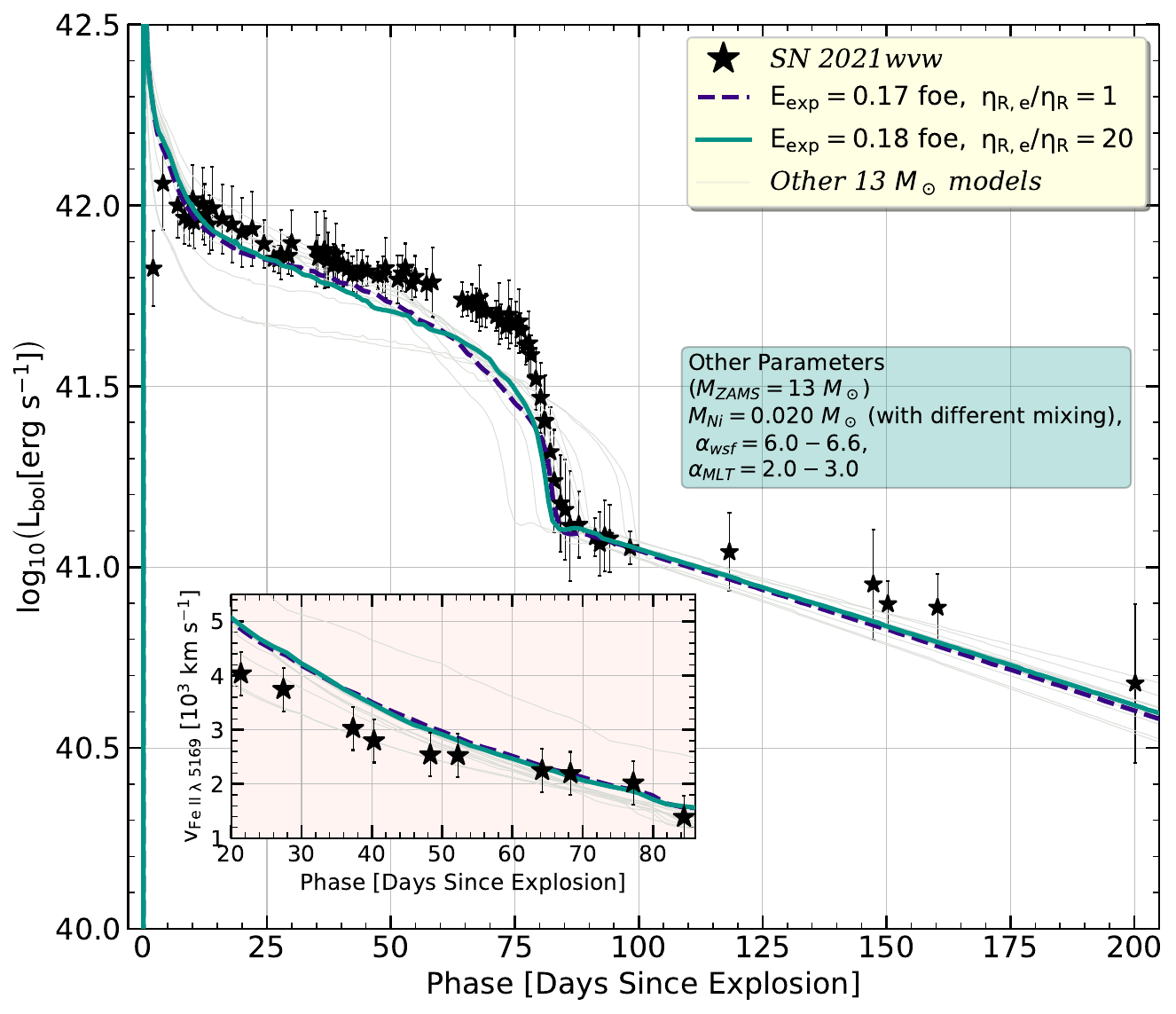}}
    \caption{Observed and modeled bolometric evolution of SN~2021wvw for $\rm 13~M_\odot$ ZAMS models with different sets of parameters. The inset in the left bottom shows the corresponding modeled and observed \ion{Fe}{2}~5169 velocities.}
    \label{fig:mesa1}
\end{figure}

\begin{figure}[htb!]
    \centering
    \resizebox{\hsize}{!}{\includegraphics{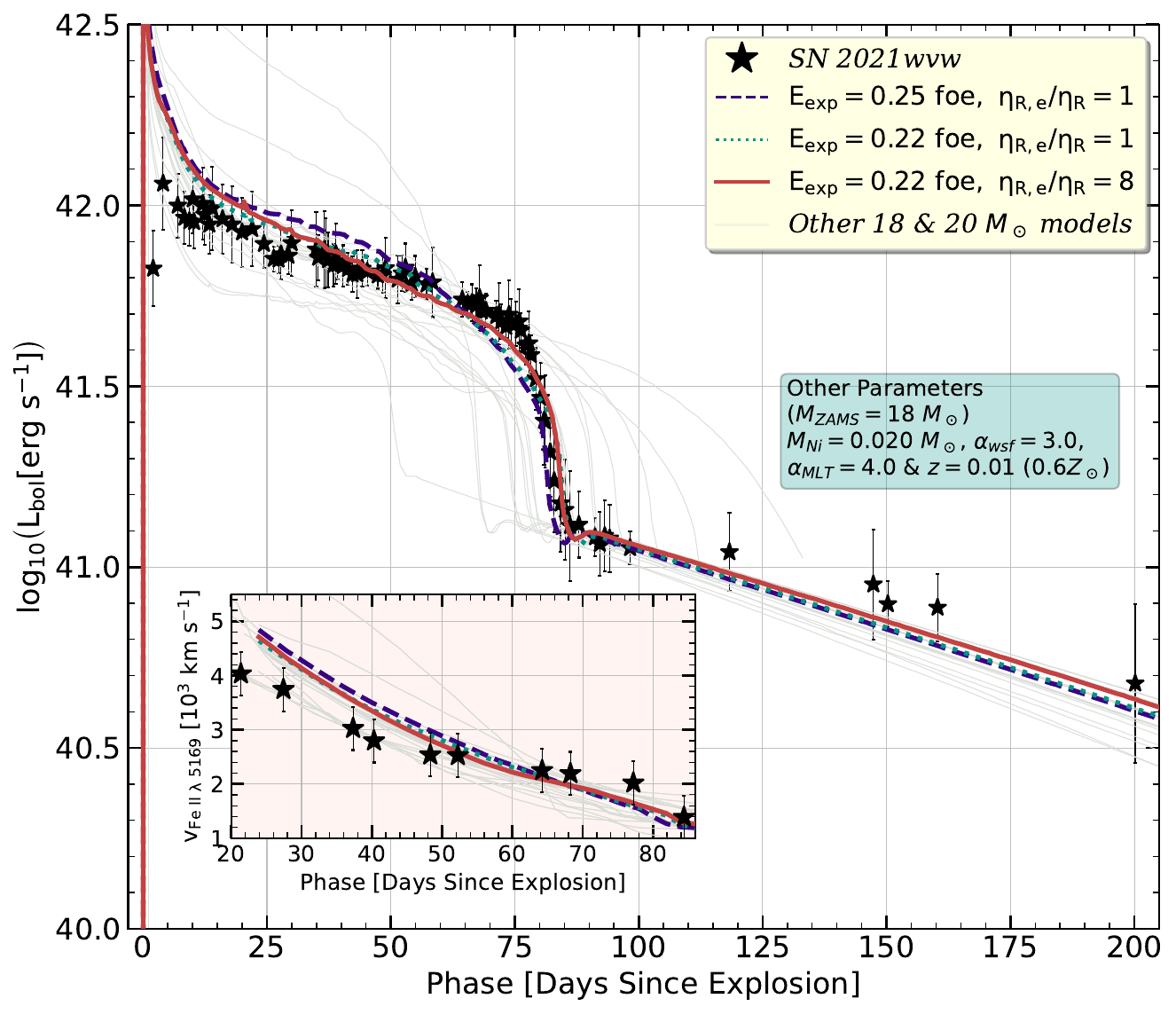}}
    \caption{Observed and modeled bolometric evolution of SN~2021wvw for $\rm 18~M_\odot$ ZAMS models with different sets of parameters. The solid red curve gives the best description of the model. The inset in the bottom left shows the corresponding modeled and observed \ion{Fe}{2}~5169 velocities.}
    \label{fig:mesa2}
\end{figure}

Firstly, we evolve 13~$\rm M_\odot$ ZAMS mass models with solar metallicity for the lower mass end. We change the wind scaling ($\alpha_{wsf}$) in steps and explode each progenitor with various explosion energies until we match the plateau luminosity and its duration. Some of the resulting bolometric light curves and corresponding \ion{Fe}{2}~5169 velocities are presented in Figure~\ref{fig:mesa1}, which are compared with the observed values. As stated earlier, we do not attempt to match the initial 10-20 days of observations exactly with models due to lack of relevant observations. We find that the velocities, plateau luminosity, and nickel tail match reasonably well for low-mass RSG models. However, these models could not reproduce the observed slow decline during the plateau phase and the sharp transition from the plateau to the tail phase. A sharp decline for SN~2005cs was obtained by increasing the strength of RTI mixing, as shown in \citep{Paxton2018}. As a more thoroughly mixed ejecta is expected to cause a steeper plateau drop due to a more even distribution of H throughout the entire ejecta, we also attempt to vary the RTI mixing via $\eta_{R,e}/\eta_R$, which directly changes the density structure as well as the abundance structure of the progenitor and the varied degree of mixing of species. Even for a value as high as $\eta_{R,e}/\eta_R=20$, we only observe slight changes in the model light curves, but not significant enough to satisfy the observed transition (refer Figure~\ref{fig:mesa1}).

We proceed further to explore and explode the higher ZAMS mass models in the range 18-20~$\rm M_\odot$ which plausibly lie on the upper mass limit for the directly detected progenitors of Type II SNe \citep{2009MNRAS.395.1409S, 2020MNRAS.496L.142D}. The resulting models are shown in Figure~\ref{fig:mesa2} with colored lines representing the best match to the observations (other models are in gray color). Owing to their large progenitor radii ($\rm \sim 1000~R_\odot$) at the mixing length $\alpha_{\rm MLT}=2$, the initial models were too bright to fit the plateau luminosities even with very low explosion energies. Hence, we evolved slightly compact progenitors to match the plateau decline and luminosities by varying the $\alpha_{MLT}$ and metallicity $z$. For $\rm \alpha_{MLT}=4.0\ \&\ z=0.6Z_\odot$, we could obtain a considerable match with the observed light curves for explosion energies of $\approx$0.22 to 0.25 foe with $\rm M_{ej}=4.7~M_\odot$. This value of $\alpha_{\rm MLT}$ is on the higher end of typically-considered values \citep[see, e.g.][]{Goldberg2020b}, and is consistent with 3D simulations of convective RSG envelopes \citep{2022ApJ...933..164Goldberg}. The transition to the end of the plateau obtained for these models is inherently sharp, which is further matched well by varying the RTI parameter. We could replicate the observed transition profile for $\eta_{R,e}/\eta_R=8$. The mass of $\rm ^{56}Ni$ required to fit the observed light curve is similar to the earlier estimates with $\rm M_{Ni}\approx0.020~M_\odot$. The ejecta mass and explosion energies obtained through hydrodynamical modeling are lower than that obtained from the semi-analytical approach. However, such discrepancies between semi-analytic and detailed modeling are fairly common in the literature (see for example, \citet{2019ApJ...876...19S,2023ApJ...954..155T}). This could be due to various simplified approximations in the semi-analytical work, including the assumed density and velocity profile of the ejecta, as well as the assumption of a simple two-zone ejecta with a grey opacity treatment independent of metallicity \citep{NagyVinko2016}.

\begin{figure}[htb!]
    \centering
    \resizebox{\hsize}{!}{\includegraphics{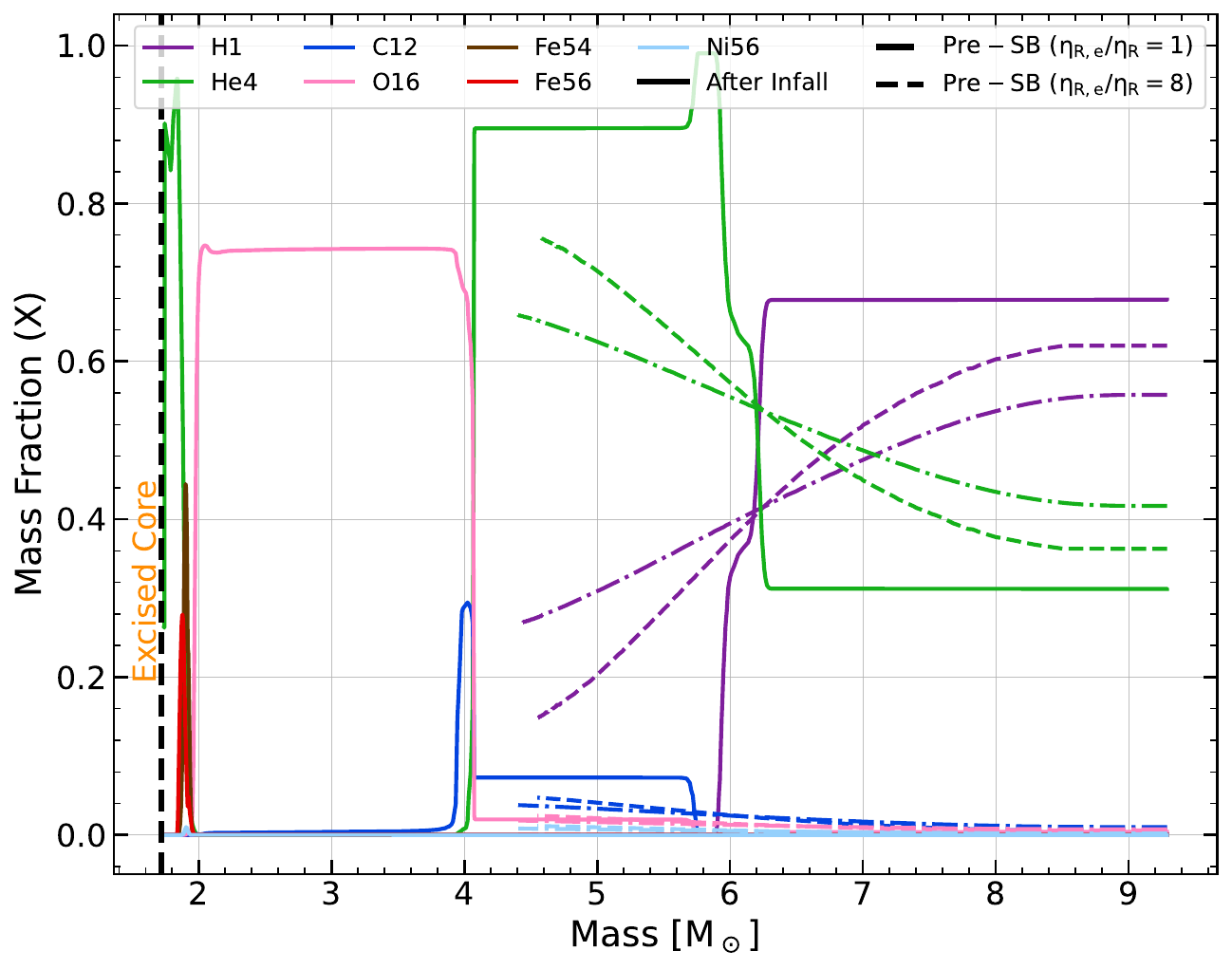}}
    \caption{MESA+STELLA structures for different cases of $\rm 18~M_\odot$ ZAMS models with different RTI parameter. A few species out of the 22 species network used in the modeling are shown here. Solid lines present the mass fraction just after we inject the explosion energy. The other two dashed lines show the final ejecta structure before the shock breakout (SB) for different $\eta_{R,e}/\eta_R$ values. The final ejecta profiles suffer from significant fallback during the shock-propagation phase, which we discuss in Section \ref{sec:fallback}.}
    \label{fig:mesa3}
\end{figure}

We show the structural differences in the various models considering the effect of the RTI parameter in Figure~\ref{fig:mesa3} using a few species out of the 22 species network used in the modeling. Solid lines represent the mass fraction just after we inject the explosion energy. The other two dashed lines show the final ejecta structure before the shock breakout (SB) for different $\eta_{R,e}/\eta_R$ values. The figure shows that the higher $\eta$ ratio weakens the RTI mixing with increasing species concentration towards the inner layers. At the boundary interface, the gradient is steeper for a higher $\eta$ ratio. Due to the small explosion energies, the models experience significant fallback during the shock-propagation phase as reverse shocks off the steep density gradients at various compositional boundaries sweep marginally-unbound material back onto the inner boundary. This is also evident in Figure~\ref{fig:mesa3}, where the inner boundary of the final pre-SB structure is at a significantly higher mass co-ordinate ($\rm \approx 4.5~M_\odot$) than what was initially excised as a core remnant mass ($\rm \approx 1.7~M_\odot$). The detailed fallback treatment in \texttt{MESA} is described in \citet{2019ApJ...879....3GGOLD}. Due to the relatively low core binding energy in the suite of 13$\rm M_\odot$ progenitors, we find only 0.2 to 0.4 $\rm M_\odot $ of material is falling onto the core in 13~$\rm M_\odot$ progenitor case, whereas it is much larger for high mass scenarios reaching up to 2-3 $\rm M_\odot$ (owing to the larger core binding energy of the high-mass progenitors). Approximately 1~$\rm M_\odot$ of fallback was also present in the SN~2005cs models \citep{Paxton2018} even for an initial low mass progenitor (13~$\rm M_\odot$).

\section{Discussion}
\label{sec:discussion}
\subsection{Scaling relation degeneracies and model differences for short-plateau SNe}

\begin{figure}[htb!]
    \centering
    \resizebox{\hsize}{!}{\includegraphics{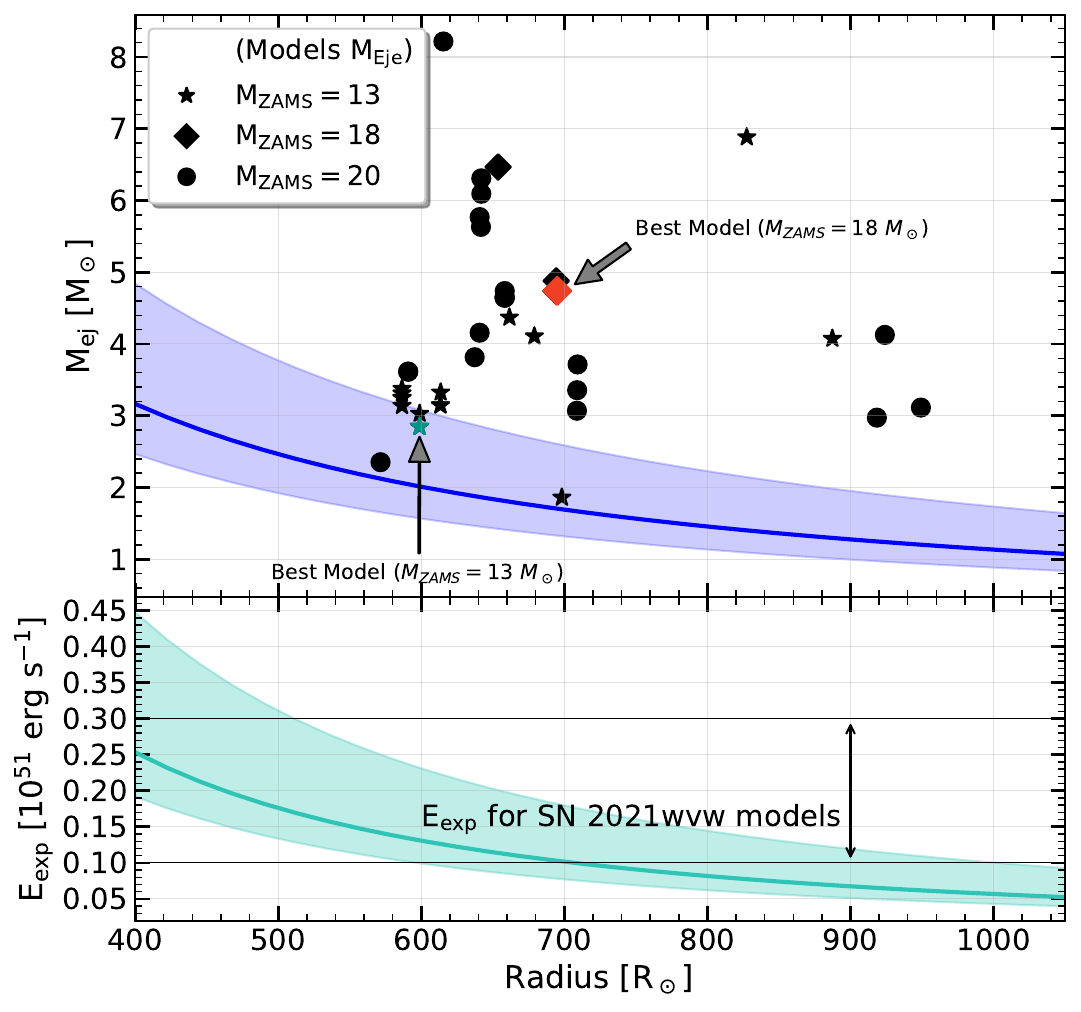}}
    \caption{ Plausible $E_{exp}$ and $M_{ej}$ ranges plotted from the scaling relations obtained in \citet{2019ApJ...879....3GGOLD}. The scatter points represent the ejecta masses obtained for various models utilized in this work. The energy values for all the evolved models are between 0.1 to 0.3 foe. The shaded regions include the values obtained considering the errors in the observables.}
    \label{fig:mesa4}
\end{figure}

Many works have highlighted the non-uniqueness of hydrodynamical modeling of SN-IIP lightcurves and plateau velocities \citep{2019ApJ...879....3GGOLD,Dessart2019,Martinez2019,Goldberg2020b}. Semi-analytical scalings between luminosity and plateau duration with progenitor properties thus entail families of explosions which may produce qualitatively similar lightcurves, with higher $\rm M_{ej}$ and $\rm E_{exp}$ at lower $R$ being comparable to smaller $\rm M_{ej}$ and $\rm E_{exp}$ at higher $R$ \citep{Popov1993,Kasen2009,2016Sukhbold,2019ApJ...879....3GGOLD,Goldberg2020b}. 
We compare a selection of our \texttt{MESA} models (from Section~\ref{sec:progenitor}) to the scaling relations obtained by \citet{2019ApJ...879....3GGOLD} to estimate a comprehensive set of ejecta mass and explosion energies, shown in Fig~\ref{fig:mesa4}. We note that these scaling relations were calibrated to higher Ni masses and more typical (i.e., less-stripped) events. We do not take these scaling relations as the absolute truth in this regime, but rather, show them as representative of the degeneracies characteristic of SNe IIP \citep{Dessart2019,2019ApJ...879....3GGOLD,Goldberg2020b}, and use them to motivate and contextualize our hydrodynamical modeling efforts.
For radii between 400-1000~$\rm R_\odot$, we find the explosion energy varies from $\rm \approx2.5\times10^{50}~erg~s^{-1}$ to much lower $\rm 5\times10^{49}~erg~s^{-1}$.  

For the given radii range, the predicted ejecta masses are less than 3~$\rm M_\odot$.  The modeled ejecta masses lie somewhat above the values obtained utilizing scaling relations for all the progenitors, possibly due to the smaller ratio of core mass to envelope mass in the sample used to calibrate the scalings compared to the models presented here. The explosion energy provides good matching values. These relations tend to give similar values obtained by semi-analytical modeling for the much more compact radii ($\rm<400~R_\odot$), also seen in the case of another short plateau SN~2018gj \citep{2023ApJ...954..155T}.  

In both the low and high mass cases for SN~2021wvw, we find apparent differences in the early phase ($\rm <40~d$) modeled and observed velocities. The differences are significant in the $\rm 13~M_\odot$ models. This tension is further increased in low-mass models when we try to match the observed plateau luminosity by increasing their progenitor radius. In other modeling works, it has been noted that the \texttt{MESA+STELLA} models provide an excellent velocity match with typical Type IIP SNe observed velocities from the early phase until the photospheric phase, which is not the case for the short plateau events.

\subsection{Fallback during the shock propagation phase\label{sec:fallback}}

In a majority of the modeled sub-luminous SNe that are the result of low-energy explosions, whether they come from low to moderate mass (8-18~$\rm M_\odot$) RSGs \citep{2000A&A...354..557C, 2017MNRAS.464.3013P, 2018MNRAS.473.3863L, 2022MNRAS.513.4983V} or high-mass RSG explosions \citep[$\rm >20~M_\odot$][]{2003MNRAS.338..711Z}, there are discussions related to fallback material onto the core. 
Namely, when the total explosion energy is positive but only comparable in magnitude to the total binding energy of the progenitor star, late-time fallback from reverse shocks during the pre-SBO phase may sweep marginally unbound material back onto the central remnant (\citealt[see, e.g.][]{Colgate1971,Perna2014}. 
In some cases, the central remnant has been speculated to turn into a black hole post-accretion, but with no observational evidence \citep{2003MNRAS.338..711Z}. In other cases, very late-time enhanced luminosity is associated with the accretion of material to the central remnant \citep{2020MNRAS.496...95G}. For many of these objects, the $\rm ^{56}Ni$ mass obtained is an order of magnitude or even much lesser than the $\rm ^{56}Ni$ mass obtained for SN~2021wvw. Further, the velocity obtained for these cases is much less than the usual Type II expansion velocities.

Interestingly, the short plateau and a sharp transition to the plateau phase are remarkable features for SN~2021wvw, which are unusual for low to intermediate luminosity SNe. Given the low inferred $\rm E_{exp}$, the short plateau length requires a low H-rich ejecta mass for both low-mass and high-mass progenitors, which could be the result of a higher mass loss during evolution. Such high mass loss might be consistent with the notion that the sharp drop from the plateau is actually \textit{excess} luminosity during the plateau drop driven by late-time interaction with previously ejected material.
But, as observed in the spectral evolution (Section~ \ref{fig:spectraevolution}), there are no discernible CSM signatures in the spectra. On the other hand, if there is an actual fallback (as occurs during hydrodynamical modeling in Section~\ref{sec:progenitor}) of the inner layers onto the core, the inward receding photosphere may reach earlier to the base of the H-rich ejecta, giving a short plateau with a sharp transition. 
This may manifest in late-time signatures of accretion if such accretion persists \citep[see, e.g.][]{Dexter2013,Moriya2019}. However, the lack of late time light curve (beyond 300 d) and spectral information restricts us from saying anything about further observational signatures of fallback accretion. 

While the short plateau and its sharp transition could be due to fallback, further discussion of the physical consequences of this fallback and ascertaining its influence on the sharp transition from plateau requires further detailed modeling, which is beyond the scope of this work. 
We nonetheless encourage follow-up observations searching for any signatures of continued accretion or very late-time circumstellar interaction from this unique event. 

\subsection{SN~2021wvw in the Type II domain}

\begin{figure}[htb!]
    \centering
    \resizebox{\hsize}{!}{\includegraphics{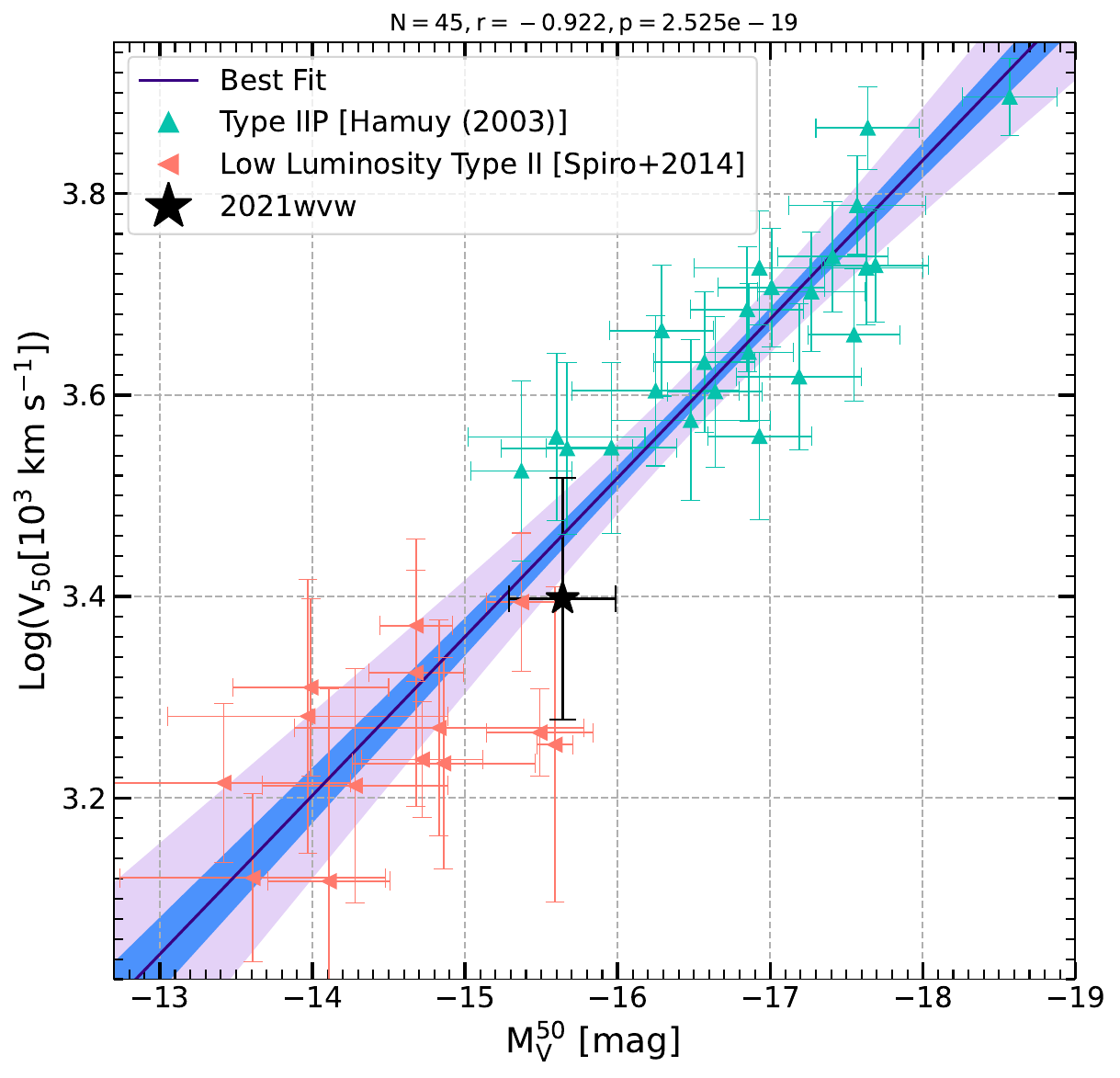}}
     \resizebox{\hsize}{!}
    {\includegraphics{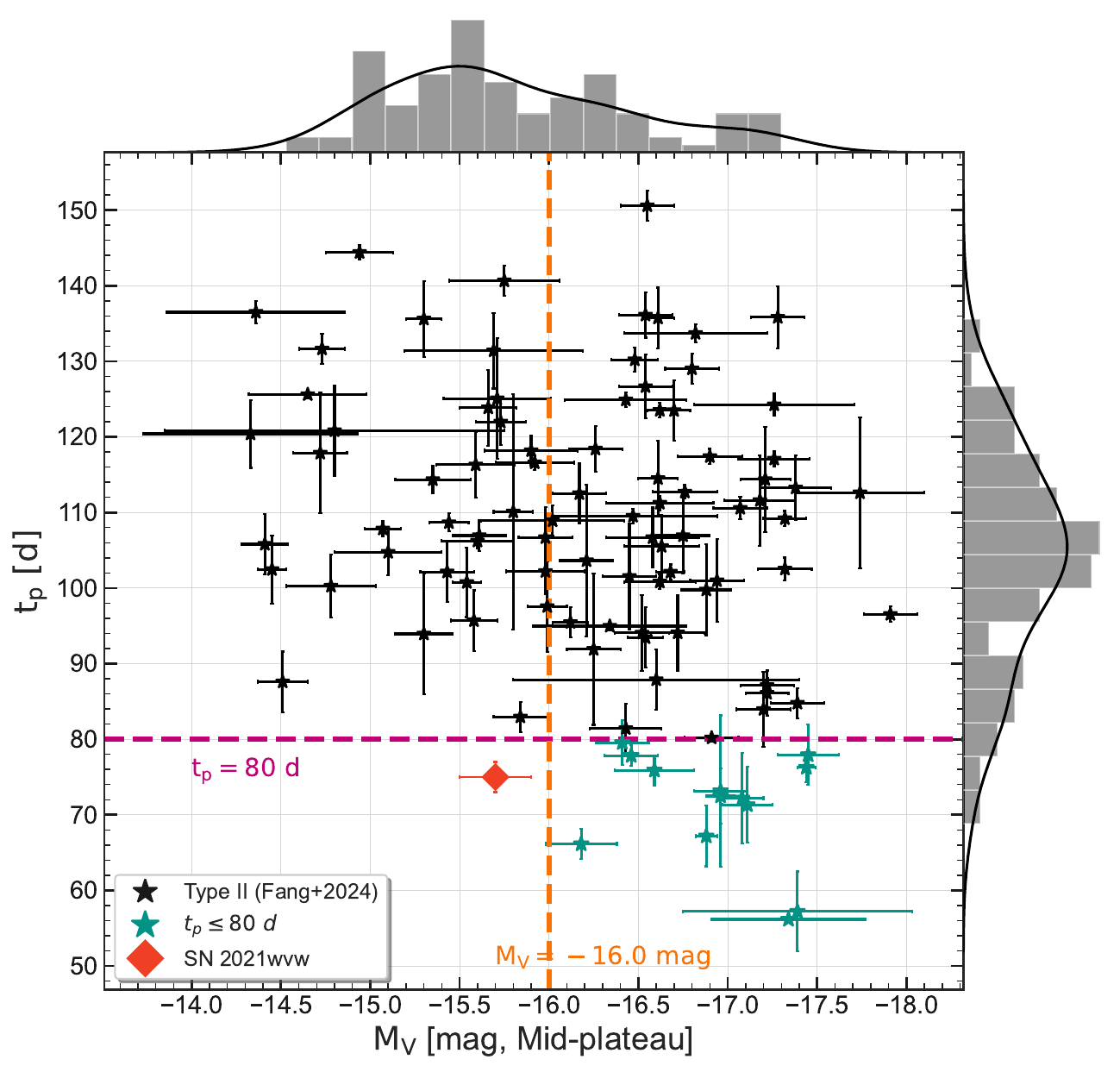}}
    \caption{\textit{Top:} Correlation between plateau brightness at 50~d, $\rm M_V^{50}$ and expansion velocities at 50~d after explosion. \textit{Bottom:} Mid-plateau brightness, $M_V$ versus plateau duration ($\rm t_p$) for a large sample including a wide range of Type II SNe obtained from \citet{2024arXiv240401776F}. }
    \label{fig:corr}
\end{figure}

 We compare SN~2021wvw with a large sample of normal Type IIP SNe \citep{2003ApJ...582..905H} and low-luminosity Type II SNe \citep{2014MNRAS.439.2873S} as shown in Fig~\ref{fig:corr}. SN~2021wvw fits well in the established tight correlation between expansion velocity and luminosity for Type II SNe at 50~d. Moreover, we find it bifurcating the two populations in both luminosity and expansion velocities. In this space, it is a bridging object between the normal Type IIP SNe and under luminous ones. Apart from this expected behavior, SN~2021wvw is unique due to its short plateau and low luminosity. Considering existing works  (e.g., refer Fig~17 in \citealt{2016MNRAS.459.3939Valenti}) showing a correlation between plateau luminosity and plateau duration, SN~2021wvw clearly is an outlier. Even for a larger sample for all Type II subclasses \citep{2024arXiv240401776F}, SN~2021wvw stands apart, as is evident in the bottom panel of Fig~\ref{fig:corr}. SN~2021wvw has the shortest plateau among all the intermediate and low-luminosity SNe. In contrast, it is the faintest SN among all the short plateau subclass of Type IIP SNe presented in the sample and, presumably, in the literature.  

\section{Summary}
\label{sec:summary}
This work provides a comprehensive set of multi-band photometric and optical spectroscopic observations of an under-luminous, short-plateau supernova SN~2021wvw. We have presented detailed light curves and spectral comparisons with other short-plateau SNe. The light curves and spectra are modeled to obtain the physical parameters of the explosion. Some of the key findings are summarized as follows:

\begin{itemize}
    \item SN~2021wvw is fainter (at $\rm M_r\approx-16~mag$) compared with other short-plateau SNe and shows the shortest plateau ($\rm \approx 75~d$) among the intermediate luminosity SNe, with a sharp transition period of $\sim$~10~d from plateau to tail phase.
    \item The ejecta expansion velocities are slowly evolving and lie below the 1-$\sigma$ lower bound compared to a large sample of Type II SNe.
    \item Early spectra show fewer metallic features as compared to other short-plateau and sub-luminous SNe. The lack of metal features is evident till the last spectrum (+95~d) presented here.
    \item Detailed \texttt{MESA+STELLA} hydrodynamical modeling disfavors the lower mass RSG models and is more inclined towards the higher mass end of RSGs. A compact progenitor with 18~$\rm M_\odot$ ZAMS mass, radius of 650-700~$R_\odot$ and a final H-rich ejecta mass of $\rm \approx 5~M_\odot$ is seen to provide a good fit to the observed properties.
    \item Modeling also suggests a low explosion energy ($\rm \approx 0.23\times10^{51} erg$) with an estimated $\rm 0.020~M_\odot$ of radioactive $\rm ^{56}Ni$.
\end{itemize}

With the increasing number of short-plateau SNe, we find that these events have varied luminosities, synthesized $^{56}$Ni masses, and expansion velocities. It is evident that these are not restricted to moderate to luminous events, as seen previously. With the upcoming large surveys such as LSST, this number would only increase and possibly make the Type IIP class or subclasses more homogenous in different parameter spaces. 


\section{Software and third party data repository citations} \label{sec:cite}

\vspace{5mm}
\facilities{HCT: 2-m, GIT: 0.7-m, ZTF, ATLAS}

\software{astropy \citep{astropy:2013, astropy:2018, astropy:2022},
ds9 \citep{2000ascl.soft03002S},
emcee \citep{2013PASP..125..306F},
IRAF \citep{1993ASPC...52..173T},
Jupyter-notebook \citep{jupyter},
matplotlib \citep{Hunter:2007},
MESA,
numpy \citep{harris2020array},
pandas \citep{mckinney-proc-scipy-2010, reback2020pandas},
plot\_atlas\_fp.py \citep{Youngfp} 
scipy \citep{2020SciPy-NMeth},
STELLA,
SYNAPPS}

\section*{Acknowledgements}
We are grateful to the anonymous referee for thoroughly evaluating the manuscript, which helped improve it.

RST would like to acknowledge Dr. Takashi J. Moriya for his insights and helpful discussions on this supernova.

RST and JAG thank Dr. Daichi Hiramatsu for readily providing observational data for a few short-plateau SNe, as well as for valuable discussions.

DKS acknowledges the support provided by DST-JSPS under grant number DST/INT/JSPS/P 363/2022.

GCA thanks the Indian National Science Academy for support under the INSA Senior Scientist Programme.

The Flatiron Institute is supported by the Simons Foundation. 

The GROWTH India Telescope (GIT) is a 70-cm telescope with a 0.7-degree field of view, set up by the Indian Institute of Astrophysics (IIA) and the Indian Institute of Technology Bombay (IITB) with funding from Indo-US Science and Technology Forum and the Science and Engineering Research Board, Department of Science and Technology, Government of India. It is located at the Indian Astronomical Observatory (IAO, Hanle). We acknowledge funding by the IITB alumni batch of 1994, which partially supports the operation of the telescope.

We thank the staff of IAO, Hanle, CREST, and Hosakote, who made these observations possible. The facilities at IAO and CREST are operated by the Indian Institute of Astrophysics, Bangalore. 

This research has made use of the High Performance Computing (HPC) resources
\url{https://www.iiap.res.in/?q=facilities/computing/nova} made available by the Computer Center of the Indian Institute of Astrophysics, Bangalore.

This work has made use of data from the Asteroid Terrestrial-impact Last Alert System (ATLAS) project. The Asteroid Terrestrial-impact Last Alert System (ATLAS) project is primarily funded to search for near earth asteroids through NASA grants NN12AR55G, 80NSSC18K0284, and 80NSSC18K1575; byproducts of the NEO search include images and catalogs from the survey area. This work was partially funded by Kepler/K2 grant J1944/80NSSC19K0112 and HST GO-15889, and STFC grants ST/T000198/1 and ST/S006109/1. The ATLAS science products have been made possible through the contributions of the University of Hawaii Institute for Astronomy, the Queen’s University Belfast, the Space Telescope Science Institute, the South African Astronomical Observatory, and The Millennium Institute of Astrophysics (MAS), Chile.

This research has made use of the NASA/IPAC Extragalactic Database (NED), which is funded by the National Aeronautics and Space Administration and operated by the California Institute of Technology.

This research has made use of the NASA/IPAC Infrared Science Archive, which is funded by the National Aeronautics and Space Administration and operated by the California Institute of Technology

\appendix

\section{DATA}

The following section provides the apparent magnitudes obtained for SN~2021wvw. The magnitudes are given in Table~\ref{app-tab:mags}.

\startlongtable
\begin{deluxetable*}{cccccccc}
\label{app-tab:mags}
\tabletypesize{ \scriptsize}
\tablecaption{Photometric observations of SN~2021wvw from GIT and HCT.}
\tablehead{
   \colhead{JD (2459000+)}         &  \colhead{Phase$^\dagger$ (d)}         & \colhead{g (mag)}    &  \colhead{V (mag) }&  \colhead{r (mag)} &  \colhead{R (mag)} & \colhead{i (mag)} &  \colhead{z (mag)}      }
\startdata
458.3	&	8.4	&		18.07	 $\pm$ 	0.18	&	-	&	17.68	 $\pm$ 	0.09	&	-	&	17.72	 $\pm$ 	0.11	&	-	\\
459.3	&	9.4	&		18.10	 $\pm$ 	0.12	&	-	&	17.68	 $\pm$ 	0.08	&	-	&	17.65	 $\pm$ 	0.11	&	17.45	 $\pm$ 	0.11	\\
460.2	&	10.3	&		18.11	 $\pm$ 	0.15	&	-	&	17.65	 $\pm$ 	0.10	&	-	&	17.64	 $\pm$ 	0.14	&	-	\\
462.3	&	12.4	&		-	&	-	&	17.66	 $\pm$ 	0.05	&	-	&	17.63	 $\pm$ 	0.06	&	-	\\
463.3	&	13.4	&		18.12	 $\pm$ 	0.10	&	-	&	17.67	 $\pm$ 	0.05	&	-	&	-	&	17.36	 $\pm$ 	0.13	\\
465.3	&	15.4	&		18.17	 $\pm$ 	0.10	&	-	&	-	&	-	&	-	&	-	\\
471.3	&	21.4	&		-	&	-	&	-	&	-	&	-	&	17.40	 $\pm$ 	0.18	\\
474.3	&	24.4	&		18.30	 $\pm$ 	0.16	&	-	&	17.69	 $\pm$ 	0.08	&	-	&	17.66	 $\pm$ 	0.07	&	17.44	 $\pm$ 	0.10	\\
476.3	&	26.4	&		18.40	 $\pm$ 	0.12	&	18.06	 $\pm$ 	0.01	&	17.81	 $\pm$ 	0.05	&	17.56	 $\pm$ 	0.01	&	17.64	 $\pm$ 	0.06	&	-	\\
477.3	&	27.4	&		-	&	18.05	 $\pm$ 	0.01	&	17.79	 $\pm$ 	0.06	&	17.55	 $\pm$ 	0.02	&	17.64	 $\pm$ 	0.07	&	17.49	 $\pm$ 	0.09	\\
478.2	&	28.3	&		-	&	-	&	-	&	-	&	17.62	 $\pm$ 	0.12	&	-	\\
479.3	&	29.4	&		-	&	-	&	17.80	 $\pm$ 	0.06	&	-	&	17.70	 $\pm$ 	0.08	&	-	\\
485.3	&	35.4	&		-	&	-	&	17.76	 $\pm$ 	0.05	&	-	&	17.66	 $\pm$ 	0.06	&	-	\\
486.3	&	36.4	&		18.49	 $\pm$ 	0.11	&	-	&	17.65	 $\pm$ 	0.05	&	-	&	17.63	 $\pm$ 	0.08	&	17.56	 $\pm$ 	0.12	\\
487.3	&	37.4	&		-	&	-	&	17.75	 $\pm$ 	0.06	&	-	&	17.69	 $\pm$ 	0.07	&	17.51	 $\pm$ 	0.11	\\
488.4	&	38.5	&		18.53	 $\pm$ 	0.15	&	-	&	17.75	 $\pm$ 	0.09	&	-	&	17.67	 $\pm$ 	0.08	&	-	\\
489.2	&	39.3	&		18.51	 $\pm$ 	0.14	&	-	&	17.76	 $\pm$ 	0.09	&	-	&	17.61	 $\pm$ 	0.11	&	-	\\
490.2	&	40.3	&		18.55	 $\pm$ 	0.14	&	18.07	 $\pm$ 	0.01	&	17.78	 $\pm$ 	0.06	&	17.56	 $\pm$ 	0.01	&	17.66	 $\pm$ 	0.09	&	-	\\
491.3	&	41.4	&		18.58	 $\pm$ 	0.16	&	-	&	17.79	 $\pm$ 	0.08	&	-	&	17.61	 $\pm$ 	0.09	&	17.52	 $\pm$ 	0.11	\\
492.2	&	42.3	&		18.62	 $\pm$ 	0.14	&	-	&	17.81	 $\pm$ 	0.11	&	-	&	17.68	 $\pm$ 	0.10	&	17.46	 $\pm$ 	0.11	\\
493.3	&	43.4	&		18.62	 $\pm$ 	0.15	&	-	&	17.78	 $\pm$ 	0.09	&	-	&	17.72	 $\pm$ 	0.09	&	17.55	 $\pm$ 	0.12	\\
494.2	&	44.3	&		18.54	 $\pm$ 	0.14	&	-	&	17.78	 $\pm$ 	0.09	&	-	&	17.66	 $\pm$ 	0.14	&	-	\\
495.4	&	45.5	&		18.57	 $\pm$ 	0.15	&	-	&	17.81	 $\pm$ 	0.09	&	-	&	17.64	 $\pm$ 	0.11	&	17.56	 $\pm$ 	0.14	\\
497.3	&	47.4	&		18.60	 $\pm$ 	0.15	&	-	&	17.83	 $\pm$ 	0.12	&	-	&	17.67	 $\pm$ 	0.13	&	17.53	 $\pm$ 	0.16	\\
498.2	&	48.3	&		18.58	 $\pm$ 	0.12	&	-	&	17.81	 $\pm$ 	0.12	&	-	&	17.72	 $\pm$ 	0.09	&	-	\\
501.3	&	51.4	&		18.72	 $\pm$ 	0.20	&	-	&	17.77	 $\pm$ 	0.12	&	-	&	-	&	17.61	 $\pm$ 	0.14	\\
502.2	&	52.3	&		18.60	 $\pm$ 	0.14	&	-	&	17.84	 $\pm$ 	0.07	&	-	&	17.67	 $\pm$ 	0.08	&	-	\\
503.2	&	53.3	&		18.51	 $\pm$ 	0.24	&	-	&	-	&	-	&	-	&	-	\\
504.1	&	54.2	&		18.75	 $\pm$ 	0.13	&	-	&	17.92	 $\pm$ 	0.06	&	-	&	17.66	 $\pm$ 	0.07	&	17.56	 $\pm$ 	0.08	\\
507.1	&	57.2	&		-	&	-	&	17.89	 $\pm$ 	0.08	&	-	&	17.63	 $\pm$ 	0.10	&	17.60	 $\pm$ 	0.12	\\
508.3	&	58.4	&		-	&	-	&	17.75	 $\pm$ 	0.08	&	-	&	17.73	 $\pm$ 	0.10	&	17.66	 $\pm$ 	0.15	\\
514.4	&	64.5	&		-	&	-	&	17.94	 $\pm$ 	0.13	&	-	&	-	&	17.69	 $\pm$ 	0.14	\\
515.4	&	65.5	&		18.83	 $\pm$ 	0.16	&	-	&	18.05	 $\pm$ 	0.10	&	-	&	17.83	 $\pm$ 	0.11	&	-	\\
516.3	&	66.4	&		-	&	-	&	17.93	 $\pm$ 	0.14	&	-	&	-	&	17.65	 $\pm$ 	0.21	\\
517.2	&	67.3	&		-	&	-	&	18.04	 $\pm$ 	0.07	&	-	&	17.87	 $\pm$ 	0.08	&	-	\\
518.3	&	68.4	&		18.89	 $\pm$ 	0.23	&	-	&	18.01	 $\pm$ 	0.16	&	-	&	-	&	-	\\
519.1	&	69.2	&		-	&	-	&	18.06	 $\pm$ 	0.08	&	-	&	17.89	 $\pm$ 	0.08	&	-	\\
521.2	&	71.3	&		18.85	 $\pm$ 	0.19	&	-	&	18.13	 $\pm$ 	0.11	&	-	&	-	&	-	\\
522.2	&	72.3	&		18.94	 $\pm$ 	0.20	&	-	&	18.13	 $\pm$ 	0.11	&	-	&	17.98	 $\pm$ 	0.11	&	17.72	 $\pm$ 	0.13	\\
523.2	&	73.3	&		-	&	-	&	18.21	 $\pm$ 	0.13	&	-	&	-	&	-	\\
524.2	&	74.3	&		18.99	 $\pm$ 	0.16	&	-	&	18.12	 $\pm$ 	0.07	&	-	&	17.99	 $\pm$ 	0.10	&	17.81	 $\pm$ 	0.12	\\
525.1	&	75.2	&		19.02	 $\pm$ 	0.18	&	-	&	18.13	 $\pm$ 	0.08	&	-	&	18.00	 $\pm$ 	0.09	&	-	\\
526.2	&	76.3	&		19.03	 $\pm$ 	0.22	&	-	&	18.21	 $\pm$ 	0.13	&	-	&	-	&	17.84	 $\pm$ 	0.16	\\
527.1	&	77.2	&		19.28	 $\pm$ 	0.19	&	-	&	18.29	 $\pm$ 	0.10	&	-	&	-	&	-	\\
528.2	&	78.3	&		19.38	 $\pm$ 	0.24	&	-	&	18.29	 $\pm$ 	0.13	&	-	&	18.12	 $\pm$ 	0.14	&	17.94	 $\pm$ 	0.16	\\
529.2	&	79.3	&		-	&	-	&	18.45	 $\pm$ 	0.14	&	-	&	18.18	 $\pm$ 	0.14	&	18.33	 $\pm$ 	0.20	\\
530.1	&	80.2	&		19.70	 $\pm$ 	0.16	&	-	&	18.57	 $\pm$ 	0.09	&	-	&	18.35	 $\pm$ 	0.08	&	-	\\
531.1	&	81.2	&		19.90	 $\pm$ 	0.11	&	-	&	18.79	 $\pm$ 	0.04	&	-	&	18.68	 $\pm$ 	0.06	&	-	\\
532.1	&	82.2	&		-	&	-	&	19.04	 $\pm$ 	0.05	&	-	&	-	&	-	\\
533.1	&	83.2	&		-	&	-	&	-	&	-	&	18.86	 $\pm$ 	0.07	&	-	\\
534.1	&	84.2	&		-	&	-	&	19.38	 $\pm$ 	0.14	&	-	&	19.31	 $\pm$ 	0.11	&	-	\\
535.1	&	85.2	&		-	&	-	&	19.52	 $\pm$ 	0.05	&	-	&	19.42	 $\pm$ 	0.08	&	-	\\
536.1	&	86.2	&		-	&	-	&	19.73	 $\pm$ 	0.06	&	-	&	-	&	-	\\
541.1	&	91.2	&		-	&	-	&	19.69	 $\pm$ 	0.08	&	-	&	-	&	-	\\
542.1	&	92.2	&		-	&	-	&	19.85	 $\pm$ 	0.08	&	-	&	-	&	-	\\
543.1	&	93.2	&		-	&	-	&	19.75	 $\pm$ 	0.07	&	-	&	19.50	 $\pm$ 	0.07	&	-	\\
544.1	&	94.2	&		-	&	-	&	19.77	 $\pm$ 	0.07	&	-	&	-	&	-	\\
548.2	&	98.3	&		-	&	-	&	19.78	 $\pm$ 	0.15	&	-	&	-	&	-	\\
555.1	&	105.2	&		-	&	-	&	-	&	-	&	19.59	 $\pm$ 	0.17	&	-	\\
563.1	&	113.2	&		-	&	-	&	-	&	-	&	19.64	 $\pm$ 	0.12	&	-	\\
568.2	&	118.3	&		-	&	-	&	19.67	 $\pm$ 	0.08	&	-	&	-	&	-	\\
575.0	&	125.1	&		-	&	20.95	 $\pm$ 	0.10	&	-	&	19.81	 $\pm$ 	0.05	&	-	&	-	\\
580.0	&	130.1	&		-	&	21.17	 $\pm$ 	0.28	&	-	&	-	&	-	&	-	\\
597.3	&	147.4	&		-	&	-	&	19.91	 $\pm$ 	0.10	&	-	&	-	&	-	\\
600.2	&	150.3	&		-	&	-	&	20.21	 $\pm$ 	0.19	&	-	&	-	&	-	\\
605.0	&	155.1	&		-	&	21.15	 $\pm$ 	0.26	&	-	&	20.19	 $\pm$ 	0.12	&	-	&	-	\\
610.2	&	160.3	&		-	&	-	&	20.01	 $\pm$ 	0.24	&	-	&	-	&	-	\\
625.0	&	175.1	&		-	&	21.19	 $\pm$ 	0.27	&	-	&	20.08	 $\pm$ 	0.10	&	-	&	-	\\
626.2	&	176.3	&		20.97	 $\pm$ 	0.19	&	-	&	-	&	-	&	-	&	-	\\
628.0	&	178.1	&		-	&	21.09	 $\pm$ 	0.12	&	-	&	20.44	 $\pm$ 	0.06	&	-	&	-	\\
629.2	&	179.3	&		-	&	-	&	-	&	-	&	20.50	 $\pm$ 	0.23	&	-	\\
636.0	&	186.1	&		-	&	21.03	 $\pm$ 	0.15	&	-	&	-	&	-	&	-	\\
650.1	&	200.2	&		-	&	-	&	20.63	 $\pm$ 	0.12	&	-	&	-	&	-	\\
665.0	&	215.1	&		-	&	21.65	 $\pm$ 	0.11	&	-	&	20.91	 $\pm$ 	0.10	&	-	&	-	\\
\hline
\multicolumn{4}{c}{\footnotesize {$^\dagger$Phase given for $t_{exp}=2459449.9$~JD}}
\enddata
\label{tab:wvwphot}
\end{deluxetable*}

\bibliography{SN2021wvw}{}
\bibliographystyle{aasjournal}

\end{document}